\documentclass[sigconf]{acmart}

\AtBeginDocument{%
  }

\newcommand{\update}[1]{\textbf{}\textcolor{black}{#1}}

\copyrightyear{2025}
\acmYear{2025}
\setcopyright{acmlicensed}\acmConference[CHI '25]{CHI Conference on Human Factors in Computing Systems}{April 26-May 1, 2025}{Yokohama, Japan}
\acmBooktitle{CHI Conference on Human Factors in Computing Systems (CHI '25), April 26-May 1, 2025, Yokohama, Japan}
\acmDOI{10.1145/3706598.3713997}
\acmISBN{979-8-4007-1394-1/25/04}

\usepackage{booktabs}
\usepackage{graphicx}
\usepackage{tablefootnote}

\begin{document}

\title{“Ignorance is not Bliss”: Designing Personalized Moderation to Address Ableist Hate on Social Media}


\author{Sharon Heung}
\affiliation{
  \institution{Cornell Tech}
  \city{New York}
  \state{New York}
  \country{USA}
}

\author{Lucy Jiang}
\affiliation{
  \institution{University of Washington}
  \city{Seattle}
  \state{Washington}
  \country{USA}
}

\author{Shiri Azenkot}
\affiliation{
  \institution{Cornell Tech}
  \city{New York}
  \state{New York}
  \country{USA}
}

\author{Aditya Vashistha}
\affiliation{
  \institution{Cornell University}
  \city{Ithaca}
  \state{New York}
  \country{USA}
}

\renewcommand{\shortauthors}{Heung et al.}

\begin{abstract}
Disabled people on social media often experience ableist hate and microaggressions. Prior work has shown that platform moderation often fails to remove ableist hate, leaving disabled users exposed to harmful content. This paper examines how personalized moderation can safeguard users from viewing ableist comments. During interviews and focus groups with 23 disabled social media users, we presented design probes to elicit perceptions on configuring their filters of ableist speech (e.g., intensity of ableism and types of ableism) and customizing the presentation of the ableist speech to mitigate the harm (e.g., AI rephrasing the comment and content warnings). We found that participants preferred configuring their filters through types of ableist speech and favored content warnings. We surface participants’ distrust in AI-based moderation, skepticism in AI’s accuracy, and varied tolerances in viewing ableist hate. Finally, we share design recommendations to support users’ agency, mitigate harm from hate, and promote safety.
\end{abstract}



\maketitle

\section{Introduction}
Disabled\footnote{We use identity-first language because it aligns with the language the majority of our participants used.} people experience high levels of harassment online, including ableist\footnote{\textit{Ableist} is the adjective form of the noun \textit{ableism}, used to describe discrimination and prejudice towards disabled people \cite{dunn21understanding,friedman17defining}} microaggressions \cite{heung22nothing} and hate \cite{heung24vulnerable, lyu24because, eagle23you, sannon23disability, mok23experiences, ringland19autsome}. Research has shown that experiencing ableism online can have lasting effects on disabled users’ well-being, influencing behaviors such as self-censorship \cite{heung24vulnerable} on social media. Furthermore, platform moderation often fails to effectively manage disability-related content, allowing some harmful ableist content to remain while mistakenly removing other legitimate content related to disability \cite{heung24vulnerable, lyu24because, rong22it, rauchberg22shadowbanned}. This disconnect between platform moderation and user needs highlights the potential for alternative moderation techniques to help users avoid exposure to ableist hate and harassment.

Personal moderation allows users to configure or customize some aspects of their moderation preferences, based on the content posted by other users \cite{jhaver23personalizing}. This configuration only affects the user’s view, meaning the content remains visible to other users. Currently, platforms offer some personal content moderation tools (e.g., word filters \cite{jhaver22designing}, toggles \cite{porter21today}, and sensitivity sliders \cite{instagram21sensitive}) that enable users to specify types of content they wish to avoid. Platforms also provide account-based moderation tools, letting users choose which accounts to follow and block. In addition to these platform-provided tools, several third-party applications have emerged, empowering users to personalize their content. For example, Google’s Tune Chrome Extension \cite{tune24} allows users to select levels of toxicity, and Bodyguard \cite{bodyguard24} promises that their AI system \textit{“instantly removes toxic, spam, and damaging content.”} The emerging prevalence and promise of these tools suggests that AI-based personal content moderation has the potential to help users avoid harmful and triggering content. 

While personalized moderation has potential to reduce the harms of seeing hateful content, few scholars have conducted empirical studies to gather users’ perceptions on the usability and efficacy of such tools. For example, Jhaver et al. \cite{jhaver23personalizing} gathered social media users’ perceptions on personal moderation tools based on toxicity, identifying critical needs such as clearer definitions of toxicity, more granular control options, and more transparency through examples of filtered content. We extend prior work by understanding how these tools should be designed for identity-based hate \textemdash{} specifically, ableism.

Personalized moderation systems that particularly address identity-based harms have started to emerge. For example, Intel’s Bleep \cite{porter21today} filters harmful content related to ableism, racism, and sexism. However, little is known about how effective these identity-based personal moderation tools are, whether these tools account for users' specific needs, and how these tools might be designed to better support those who may benefit from them. To our knowledge, there is no understanding of how disabled people perceive these filters that are designed to address ableism.

To address this critical gap, we examined how an ableism-specific personal moderation tool should be designed to account for the needs and preferences of disabled people. More specifically, we ask:

\textbf{RQ1:} How do disabled social media users perceive and use existing personal moderation tools for addressing ableist hate and harassment?

\textbf{RQ2:} How can personal moderation tools be designed to account for ableist speech during experiences of ableist hate and harassment?

We conducted a two-part study. First, we held initial interviews with 23 disabled social media users to introduce current personal moderation tools and gain insight on their experiences using word filters and blocking. Then, we facilitated eight 90-minute focus groups with 2-3 participants each, presenting design probes to gather feedback on new designs of AI-based personalized moderation tools for addressing ableist content. Our design probes include two ends of a personalized moderation system. First, we present various ableism-specific filter settings (e.g., binary toggle for ableism or toggle for different types of ableist speech). We then show various designs augmenting the presentation of the filtered speech (e.g., having AI rephrase the ableist comment to be less toxic or having a content warning). This contrasts with most existing filters which only fully remove the hateful comment from view. We used these design probes to elicit participants’ perceptions on why they would or would not use such tools, the desired capabilities of such tools, and AI's capability of effectively identifying and rephrasing ableist text.

We found that the majority of participants commonly addressed ableist hate and harassment by responding to the perpetrator and/or blocking certain accounts. A subset of participants had experience setting word filters (e.g., removing the r-word); however, word filters were laborious to set up and not always reliable. In response to the design probes, the majority of participants preferred configuring their filters based on types of ableist hate, as it was perceived to be more understandable regarding what types of content would be filtered. Participants also favored content warnings for empowering them to decide whether or not to view the hate. Participants also expressed concerns that AI filters would struggle to identify ableist speech, leading to the wrongful filtering of disability-related content (e.g., posts containing reclaimed words). Since personal moderation only affects an individual user’s view, participants wanted platform moderation to also adopt ableism-specific content warnings as a means of educating other users on ableism. 

Based on our findings, we discuss how personal moderation can better support the safety of users experiencing ableist hate and harassment online. We recommend for ableism-specific AI filters to move away from full removal of hateful content and instead allow for customized designs (e.g., content warnings) and nudges (e.g., notification of death threats) to support user safety and to promote user agency. We also recommend personal moderation tools to allow users to customize their filters based on specific types of ableist hate. Given the widespread distrust in platform moderation, we recommend increasing transparency and incorporating mechanisms for user control, ensuring that users can oversee and reverse filtering decisions as needed to build trust. In summary, our study contributes:
\begin{itemize}
    \item Insights into how disabled social media users perceive and utilize existing personal moderation tools (RQ1).
    \item Designs of AI-based moderation and perceptions of disabled people on AI’s capabilities of identifying ableist text (RQ2).
    \item Design recommendations for how personal moderation can be designed to address ableist speech and support user agency and safety during experiences of online hate (RQ2).
\end{itemize}
\section{Related Work}
Like other "isms" (e.g., racism and sexism), ableism is discrimination towards a social group, specifically disabled people \cite{dunn21understanding, friedman17defining}. Disability studies scholars have investigated how ableism surfaces societal perceptions of disability. For example, Campbell describes ableism to cast disability as a diminished state of being human \cite{campbell08refusing}, which leads to the compulsory preference for non disability \cite{campbell10contours}. Scholars have also understood how ableism is connected to ideals and attributes that are valued or not valued \cite{wolbring08politics}. It is ableist to assert preference for a child to read print rather than Braille or to walk rather than use a wheelchair, which is harmful for students receiving disability accommodations in school \cite{hehir07confronting}. More broadly, scholars have described ableism as a capitalistic ideology of assigning value to people's productivity. Talila A. Lewis, a disability activist and lawyer, defines ableism as a “system of assigning value to people’s bodies and minds based on societally constructed ideas of normalcy, productivity, desirability, intelligence, excellence, and fitness" \cite{lewis22working}. We draw on prior work, using ableism as a term to address the collective discriminatory experience the disability community face \textit{online} with a specific focus on \textit{ableist text.}

In this section, we first review HCI literature on ableist speech within social media. Then, we situate our work within the broader context of online moderation and personal moderation, especially with regards to the experiences of people with marginalized identities.

\subsection{Ableist Speech on Social Media}
Prior work has categorized the diverse ways in which the disability community encounters ableist hate and harassment online, through public comments to private messages\cite{sannon23disability, lyu24because, heung24vulnerable, kaur24challenges, borgos-rodriguez21understanding, ringland19autsome, eagle23you}. Sannon et al. \cite{sannon23disability} found that disability activists often experienced invalidating comments on their disability, sexual harassment, fetishization, and coordinated attacks in response to their advocacy work. Building on this work, Heung et al. \cite{heung24vulnerable} developed a taxonomy of ableist hate found across 50 disabled content creators with varying disability identities. This taxonomy includes five broader categories of ableist hate (i.e., Slurs \& Derogatory language, Violent \& Eugenics-related speech, Questioning Ability \& Denying Access, Mocking \& Invalidating Disability Identity, Objectifying the Disabled Body) and 11 specific types of ableist hate (e.g., Short Slurs; Using Disability as an Insult; Death Threats, Suicide, and Self-harm). The researchers also explored how creators' intersectional identities (e.g., race and sexuality) impacted the frequency of ableist hate, finding that LGBTQ creators experience significantly more ableist hate than non-LGBTQ creators. Heung et al. also alluded to potential differences in ways ableist hate is experienced depending on one's disability identity. For example, people with invisible disabilities (less visibly apparent disability) acknowledged that they may be less likely to experience overt forms of ableist hate, but more often experienced invalidating comments about their disability. Prior work has also shown that invalidation of disability is prominent in online ADHD communities \cite{eagle23you}. 

In addition to overt hate, disabled people also experience microaggressions, or subtle forms of ableist speech. For example, many receive patronizing comments (e.g., "you're so inspirational") and infantilizing remarks (e.g., "where's your mom?") \cite{heung22nothing, keller10microagggressions}. Prior work acknowledges that the distinction between microaggressions and overt hate is not always clear-cut; for instance, accusations of faking one’s disability were perceived to be both a microaggression and an act of overt ableist hate \cite{heung22nothing, heung24vulnerable}. 

Researchers have also examined the harms of microaggressions and overt hate and found that these experiences have significant impact on disabled people, including increased emotional distress, anxiety posting online, and self-censorship overtime \cite{heung24vulnerable, heung22nothing,johnson19inclusion, sannon23disability}. Although viewing hate can be harmful, previous research indicates that some users, particularly content creators, may tolerate certain forms of hate \cite{thomas22its,sannon23disability,heung24vulnerable, kaur24challenges}. For disability activists and creators, exposure to hate can inspire their activism and inform their educational content in creative ways \cite{sannon23disability,heung24vulnerable}. For example, Duval et al. \cite{duval21chasing} found TikTok videos advocating for disabled people or debunking disability stereotypes as a common form of playful content (e.g., using upbeat sound effects and humor). Also on Tiktok, Wang et al. \cite{wang23weaving} found autistic creators to create hashtags directly in response to ableism, such as \#ableismisntcute and \#ableistsuck.

We build on existing literature understanding ableism online by exploring how personal moderation can be designed to reduce the harm caused by exposure to ableist text. Our work is motivated by the failure ofplatform moderation in supporting disabled social media users during online hate, which we discuss next. 

\subsection{Platform Moderation}
Content moderation is the organized practice of screening and controlling for unwanted content, content that is deemed as "irrelevant, obscene or illegal" \cite{fiesler18reddit, scheuerman21framework}. Most common conceptions of moderation are of mechanisms deployed after an infraction occurs, also known as reactive moderation \cite{grimmelmann15virtues}. Current reactive moderation techniques include filtering or removing inappropriate content, suspending the offending users, or even recommending and curating alternate content \cite{mcgillicuddy2016controlling, gillespie18custodian}. Moderation can be done by human moderators, users themselves, or, increasingly via algorithms running AI models and toxicity classifiers. AI-based reactive moderation can take multiple forms, including automatically filtering out keywords, such as the AutoModerator bot \cite{automodreddit} on Reddit. Other techniques leverage natural language processing techniques to automatically detect toxicity (e.g., Perspective API \cite{perspectiveAPI}, AWS \cite{aws23}) or computer vision to detect violent or graphic content \cite{nieva24heres}). While these techniques do not necessarily remove the content from the platform, they often hide the detected content behind a warning \textemdash{} for example, Meta’s current warning reads, \textit{“Sensitive content: this video contains content that some people may find upsetting”} \cite{facebook24how}. 

However, despite AI’s ability to moderate at scale \cite{gillespie20content}, there are several pitfalls. AI-powered moderation systems often lack context and community-specific nuance, which is especially important since online discourse varies greatly depending on the audience, the place of communication, the speaker, and their tone. For example, Oliva et al. \cite{oliva21fighting} found that drag queen Twitter accounts were considered to have higher perceived levels of toxicity than Donald Trump and white nationalists when moderated by AI models unfamiliar with their lexicon. In addition, AI-based moderation systems are known to exhibit ableist, sexist, colonialist, and racist tendencies. For example, Shahid et al. ~\cite{shahid23decolonizing} show that Meta’s AI-based moderation has high false positive rate for users in the Global South and the underlying algorithms imbibe coloniality by centering Western norms and erasing minoritized expressions. While these are a few examples of AI’s shortcomings, this demonstrates the risk of further silencing already marginalized voices. 

Particularly for disabled people, research shows that platform moderation is often inadequate in protecting them from ableist hate and harassment. For example, disabled creators perceived that ableist hate is oftentimes not removed by the platform, despite being reported, forcing creators to manually delete hateful comments themselves or organize other users to report on their behalf \cite{heung24vulnerable, sannon23disability}. Furthermore, disabled social media users have felt wronged by moderation, such as being penalized by moderation when responding to trolls \cite{lyu24because} or by social media algorithms suppressing disability-related content \cite{lyu24because, choi22its, karizat21algorithmic, rauchberg22shadowbanned, heung24vulnerable}. Platforms not addressing ableist hate may contribute a chilling effect for the disability community, with fear of being overshadowed by hostile voices \cite{datasociety24}. Furthermore, with the ableist hate not removed, other disabled users may be less likely to post online; the Pew Research Center reported that 27\% of Americans have refrained from posting online after witnessing harassment \cite{duggan17online}. Beyond platform moderation not removing ableist hate, disabled users also exert additional labor given the inaccessibility of social media platforms more generally \cite{lyu24because,Niu24please}.

In our work, we explore how existing and imagined moderation techniques, including personalized moderation, are potentially suited to prevent viewing ableist hate and harassment in particular.

\subsection{Personalized Moderation}
Given varying norms across cultures and communities \cite{jiang21understanding, schoenebeck23online}, research shows that a one-size-fits-all approach to content moderation is insufficient to meet the diverse needs of users \cite{cresci22personalized}. Additionally, platforms have varying definitions of harassment and different corresponding platform policies, highlighting inconsistencies on defining and moderating online harassment across platforms \cite{pater16characterizations}. To maintain free speech online while mitigating harmful content, there is a growing call to move away from a centralized moderation to a user-centered approach \cite{engler22middleware, schoenebeck23online, cresci22personalized}. Essentially, what if users themselves could decide how they want their content to be moderated? 

Jhaver et al. \cite{jhaver23personalizing} define personal moderation as tools that \textit{“let users configure their preferences for the activity they want to avoid.”} It is important to note that this form of moderation only changes the configured user’s view, other social media users can still view the filtered content. Fukuyama et al. \cite{fukuyama} refers to this individualized approach to customization of content moderation as \textit{“middleware,”} imagining third-party services as adding an editorial layer between platforms and users. In this section, we describe the two types of personalized moderation, account-based moderation and content-based moderation, specifically highlighting its usage in the context of hateful content.

\subsubsection{Personal Account Moderation.} Personal account moderation tools enable users to mute or block a particular account, determining who they want to engage with online. Blocking or muting accounts means that the content from that account or creator will no longer appear on a user’s feed \cite{x24blocking, facebook24what}. Blocking is more restrictive than muting: a user can still interact with and view content from a muted account if they are on their profile, whereas blocking disallows a user from engaging with the other user in any way and is typically known to both parties. Blocking is typically on an individual basis; however, blocklists have emerged as a way for users to easily block many users at once and have been found to be effective in addressing online harassment \cite{jhaver18online}. Prior work has shown that disabled content creators leverage blocking to foster a safe space for themselves and their followers \cite{heung24vulnerable}. 

\subsubsection{Personal Content Moderation.} Personal content moderation allows users to make moderation decisions on individual posts based on their content alone, regardless of its source \cite{jhaver23personalizing}. Common personal content moderation tools include word filters or the ability to mute specific keywords \cite{jhaver22designing}. Prior work has found word filters useful for automatically removing toxic comments and removing potential doxxing attempts, which is the non-consensual release of private and personal information \cite{samermit23millions, thomas22its}.

Word filters on most platforms use rule-based automation, which can require labor of inputting words and variations of words on their own. In response to this difficulty, researchers have developed FilterBuddy \cite{jhaver22designing}, which allows creators to easily edit filters, including adding spelling variants of keywords, previewing the effects of specific word filters, importing word filter categories (e.g., Homophobia, Pejorative Terms for Women, and Anti-Black Racism), and sharing word filters with other creators. This type of individual rule-setting is a type of distributed content moderation \cite{jiang23a} where content creators have governance over enforcing local rules in the comments. Filtering keywords empowers content creators to automatically moderate their own account, and it is the only tool that enables end-users to preemptively limit harmful content on their profile. One can think of this as a reactive approach, hiding comments after it is posted. 

Beyond rule-based filters, personal content moderation tools have begun to integrate AI to identify and therefore filter certain types of content. Some platforms have incorporated these AI-based personal moderation tools. For instance, Twitch creators have Automod, an automated moderation tool that allows users to filter content based on these categories: discrimination and slurs, sexual content hostility, and profanity \cite{twitchautomod}. Instagram has incorporated sensitivity sliders \cite{instagram21sensitive} that default to “normal,” but users can choose “more” or “less” to indicate the amount of sensitive content users want to see in their timeline. Similarly, Google released an experimental Chrome browser extension, Tune, that allows users to customize how much toxicity they wish to see in comments across the internet \cite{jigsaw19tune}.

Despite growing interest in AI-based personal moderation tools, empirical work on end users’ perceptions on such tools is limited. One survey study found users to view personal moderation tools as a means for greater agency over their social media experience, and not an infringement on free speech \cite{jhaver23do}. In another study, Jhaver et al. \cite{jhaver23personalizing} investigated end users’ perceptions on personal content moderation tools that filtered content based on toxicity and identified several improvements to content moderation tools, such as increased clarity in definitions of what is hateful, more granularity in end user controls, and greater transparency in what content gets filtered. 

In this paper, we extend work on personalized moderation by capturing disabled social media users’ experiences with personal moderation tools (RQ1) and enriching knowledge on how these tools should be designed to account for ableist speech during hate and harassment (RQ2).

\section{Methods}
We conducted 23 interviews and eight focus groups to understand participants' prior moderation experiences and ideate on personalized moderation techniques.

\subsection{Participants}
Given that our work focuses on personalized moderation during and after experiences of ableist hate, we specifically selected participants who are social media users and havedisclosed their disability identity on social media. This approach, used by other scholars \cite{heung22nothing, heung24vulnerable} ensured that participants had firsthand experience with ableist hate directed at them because of their disability identity. Participants were excluded if they were not comfortable communicating in English or American Sign Language (ASL).

We used convenience sampling from a pool of prior participants who had previously agreed to be contacted for future studies, curated by the first author. We then conducted snowball sampling to recruit additional participants. We launched a short screening survey to confirm eligibility. All participants self-identified as having a disability and were 18+ years old. We asked participants to share their disability identity through Blaser et al.'s \cite{blaser20why} survey question design with options to select multiple disability identities and input an open-ended response. We recruited a diverse group of participants with varying disabilities and different types of social media usage; 14 of our participants self-identified as content creators and / or influencers. All participants were located in the US, Canada, or Europe. To protect the anonymity of creators, we share aggregated participant demographics in Table 1. 
\begin{table}[h] \small
\caption{Aggregated Participant Demographics}
\label{tab:ppt}
\begin{tabular}{@{}ll@{}}
\toprule
\multicolumn{2}{l}{\textbf{Participant Demographics}} \\ \midrule
\multicolumn{1}{|l|}{\textbf{Age}} &
  \multicolumn{1}{l|}{\begin{tabular}[c]{@{}l@{}}18-24 = 3\\ 25-34  = 13\\ 35-44 = 3\\ 45-54 = 4\end{tabular}} \\ \midrule
\multicolumn{1}{|l|}{\textbf{Gender}} &
  \multicolumn{1}{l|}{\begin{tabular}[c]{@{}l@{}}Male = 11\\ Female = 11\\ Trans Male = 1\end{tabular}} \\ \midrule
\multicolumn{1}{|l|}{\textbf{Race}} &
  \multicolumn{1}{l|}{\begin{tabular}[c]{@{}l@{}}White = 8\\ Black \& African American = 10\\ Latina = 1\\ Latin American = 1\\ Mixed race (e.g. Hispanic \& Asian) = 3\end{tabular}} \\ \midrule
\multicolumn{1}{|l|}{\textbf{Disability \update{\footnotemark}}} &
  \multicolumn{1}{l|}{\begin{tabular}[c]{@{}l@{}}Blind or low vision = 4\\ d/Deaf or hard of hearing = 3\\ Neurodivergent = 4\\ ADHD = 4\\ Autism = 7\\ Health-related disability = 10\\ Permanent / long-term disability = 10\\ Physical disability = 1\end{tabular}} \\ \midrule
\multicolumn{1}{|l|}{\textbf{Social Media Platform}} &
  \multicolumn{1}{l|}{\begin{tabular}[c]{@{}l@{}} X (previously Twitter) = 20\\ Facebook = 19\\ Instagram = 19\\ TikTok = 12\\ Linkedln = 10\\ Snapchat = 9\\ Reddit = 4\\ Twitch = 3\\ OnlyFans = 1\\ BlueSky = 1\end{tabular}} \\ \midrule
\multicolumn{1}{|l|}{\begin{tabular}[c]{@{}l@{}} \textbf{Experiences with} \\ \textbf{Types of Ableist Hate}\\ \update{~\citet{heung24vulnerable}}\end{tabular}} &
  \multicolumn{1}{l|}{\begin{tabular}[c]{@{}l@{}}Short Slurs = 16\\ Using Disability as an Insult = 19\\ Death Threats, Suicide, and Self-harm = 8\\ Violent \& Dehumanizing Speech = 14\\ Eugenics-Related = 9\\ Disability as Inability = 18\\ Denial and Stigmatization of Accessibility = 16\\ Mocking Disability = 17\\ Accusing of Faking Disability = 13\\ Attacking Physical Appearance = 13\\ Sexual Harassment \& Fetishization = 9\end{tabular}} \\ \bottomrule
\end{tabular}
\end{table}

\update{\footnotetext{Disability demographics were self-identified and not mutually exclusive, as many participants identified as having more than one disability identity.}}

\subsection{Procedure}
Eligible participants were invited to participate in (1) a 15-minute interview and (2) a 90-minute focus group, both on Zoom. Participants were compensated with a \$10 digital gift card for the interview and a \$50 gift card for the focus group. ASL interpreters also consented to participate and were compensated for their time.

\begin{figure*}[ht!]
    \centering
\includegraphics[width=.9\textwidth]{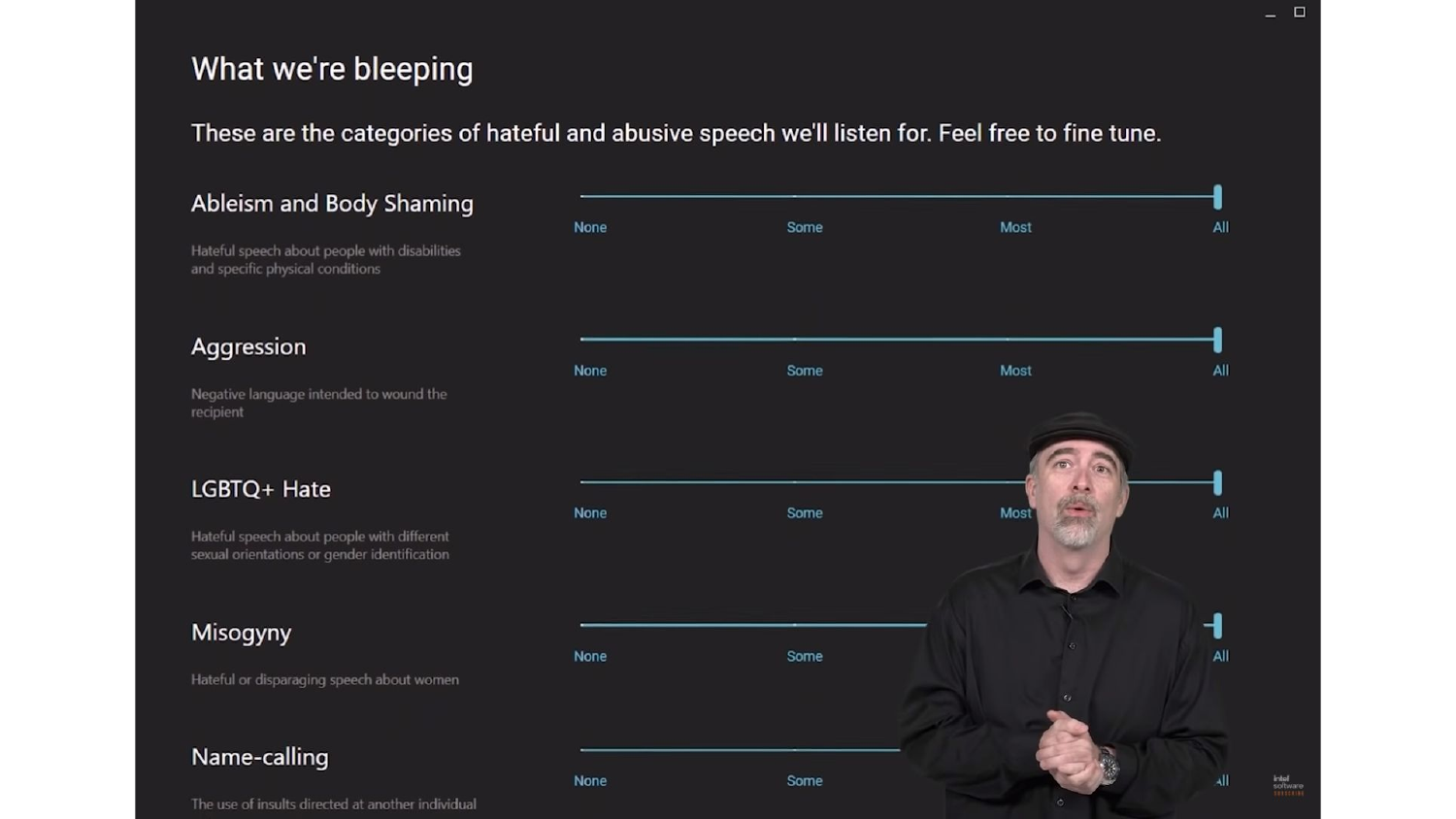}
    \caption{Screenshot of Intel's Bleep interface.}
    \label{fig:1}
    \Description{This screenshot depicts an interface titled “What we’re bleeping.” Underneath the title says “These are the categories of hateful and abusive speech we’ll’ listen for. Feel free to fine tune.” On the left hand side, there are categories to tune from and underneath each category there is a description. The categories and descriptions are as follows (from top to bottom): “Ableism and Body Shaming” underneath description is “Hateful speech about people with disabilities and specific physical conditions," “Aggression” description is “Negative language intended to wound the recipient,” “LGBTQ+ Hate” description is “Hateful speech about people with different sexual orientations or gender identification,” “Misogyny” description is “Hateful or disparaging speech about women”, and “Name-calling” description is “The use of insults directed at another individual.” On the right side of the screenshot are 5 sliders, one next to each category. Each slider with labels ranging from (left to right): "None," “Some,” “Most,” to "All." The sliders for all categories are set to the maximum (labeled "All"). In the lower right corner of the screen, there is a man with a beard and light skin tone, wearing a black shirt and hat, standing and talking in front of the interface.}
\end{figure*}

During the interviews, we asked participants about their experiences with responding to ableist hate and harassment, introduced existing personalized moderation tools, and gained insight on their experiences and challenges with such tools. Then, we discussed scheduling logistics and accessibility accommodations for the follow-up focus group study. Example of accommodation requests included a verbal description of the design probes during the focus group for blind and low vision participants and flexibility of taking breaks and turning off their video for participants with chronic conditions.

We recognize that discussing ableism may cause emotional distress. Following best practices of conducting research on online hate \cite{schafer23participatory} and a trauma-informed research approach \cite{hirsch20practicing}, we used the initial interviews to build rapport, reminded participants they could remove themselves from the study or take breaks at any point in time, and shared resources with them to cope with online hate.

After the interviews, we facilitated eight focus groups, seven of which had three participants and one had two participants. We started the focus group with introductions, expectations for creating a safe space, and preliminary feedback on Intel’s Bleep design to configure content based on ableism (see Figure \ref{fig:1}). This served as an introduction to AI-based personal moderation tools that exist today. Then we spent an hour presenting design probes, including configuring filters based on ableism (Figure 2), configuring filters based on types of ableist hate (Figure 3), and customizing the presentation of ableist hate (Figure 4). After presenting each design, we asked participants to share their thoughts about what they liked or disliked, concerns they had, and what they would have changed about the design. At the end, we asked participants to reflect on all the probes and to build their own personalized moderation tool.

The focus group setting provided participants with an opportunity to highlight their personal preferences and contrast them with others’ thoughts. We emphasized that the goal of the probes was not necessarily to understand which was “better,” but to concretely ideate on how a personalized moderation tool could be designed to mitigate the harm of viewing ableist hate. We positioned the design probes as works-in-progress requiring their expert feedback, reducing the power dynamics between researchers and participants \cite{ming21accept}. These probes also provided a starting point for participants to engage without needing to disclose personal stories in a group setting \cite{to21reducing}. 

\subsubsection{Design Probes}
Our work builds on prior research and practice in the area of personalized moderation \cite{jhaver23personalizing, haimson20trans}. For example, Jhaver et al. \cite{jhaver23personalizing} captured end users perspectives of a personalized moderation to filter out varying levels of toxicity.  Additionally, Intel's Bleep is an AI-powered tool designed to filter out identity-based hate in voice calls. While Bleep is already available for consumer use, there is limited understanding of how disabled people view such filters. We sought to explore how disabled people perceived filters specifically aimed at addressing ableism.

We used design probes to explore future designs spaces of using personalized moderation to address ableism. We shared probes related to: 1) designs to configure ableism-specific filter settings (Design Probe 1 \& 2) and 2) designs on how the tool acts on these settings; for example instead of filters fully removing the content we explore other alternatives like rephrasing the hateful comment or a content warning (Design Probe 3).

\begin{figure*}[ht!]
    \centering
    \includegraphics[width=0.7\textwidth]{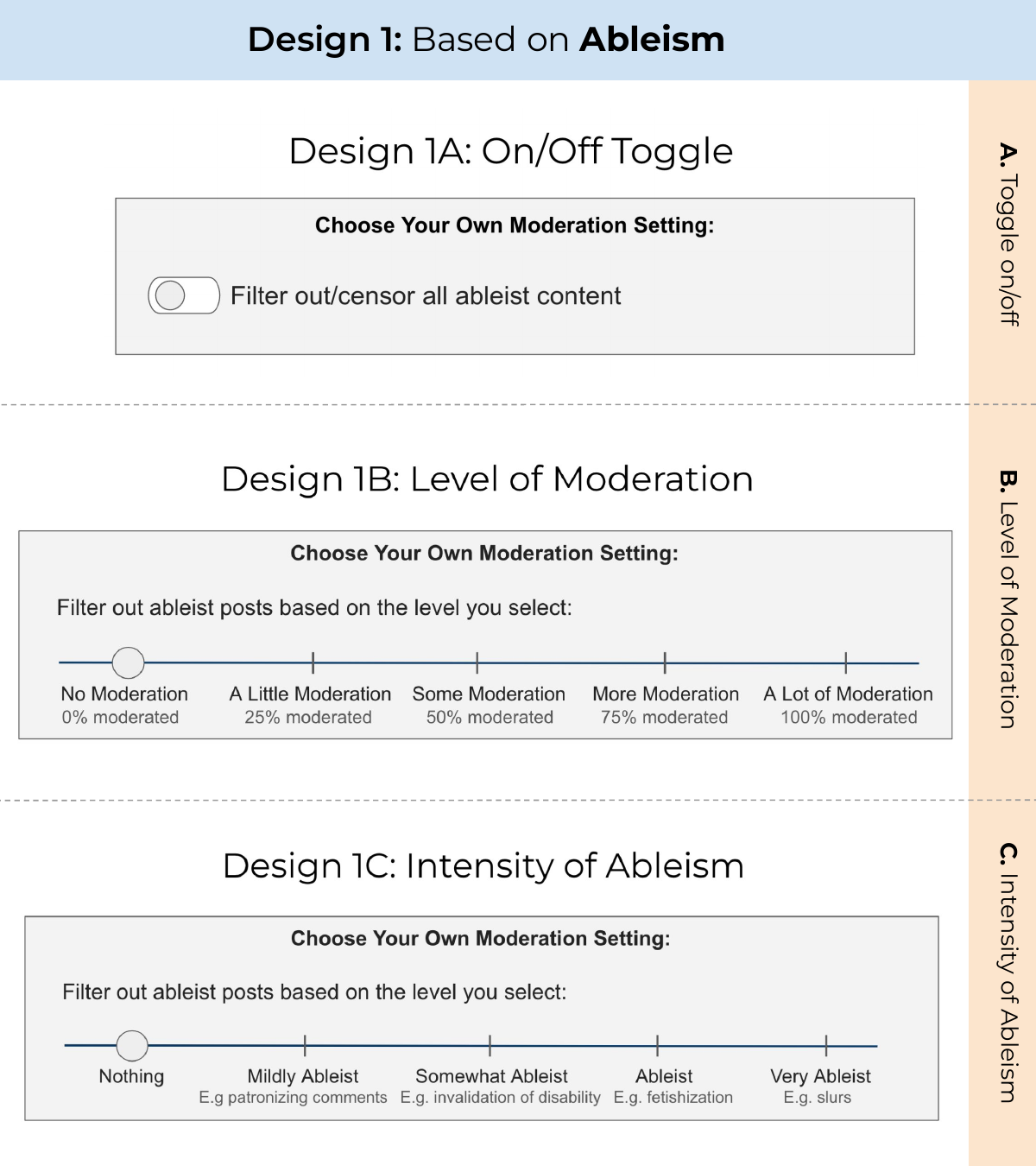}
    \caption{Diagram of Design Probe 1, configuring filters based on ableism. This includes 3 different designs: 1A (toggling ableist content), 1B (slider based on quantity of ableist posts), and 1C (slider based on intensity of ableism).}  
    \label{fig:2}
    \Description{This diagram is titled “Design 1: Based on Ableism.” There are three designs shown. On the top, there is “Design 1A: On/Off Toggle.” Underneath there is a grey box. In the box on the top, it reads "Choose Your Own Moderation Setting.” Underneath there is a simple toggle on/off switch and next to that it says “Filter out/censor all ableist content.” In the middle, there is another design titled “Design 1B: Level of Moderation.” Underneath there is another grey box. Within the grey box it says ”Filter out ableist posts based on the level you select:”. Underneath that, there is a slider that can adjusted from (left to right): "No Moderation (0\% moderated)" to "A Little Moderation (25\% moderated)", "Some Moderation (50\% moderated)", "More Moderation (75\% moderated)", and "A Lot of Moderation (100\% moderated). The slider is set to “No Moderation”. At the bottom, there is the final design titled “Design 1C: Intensity of Ableism.” There is a grey box. Inside the grey box, on the top it reads: "Choose Your Own Moderation Setting." Underneath it says ”Filter out ableist posts based on the level you select:” Below that there is a slider (from left to right) it says:"Nothing", "Mildly Ableist" E.g., patronizing comments. "Somewhat Ableist" E.g., invalidation of disability. "Ableist": E.g., fetishization of disability. "Very Ableist" E.g., slurs. The slider is set to “Nothing”. On the right side there is a yellow vertical label which categorizes the designs into three sections: A. Toggle on/off, B. Level of Moderation, and C. Intensity of Ableism. Each section aligns with the three different types of designs. }
\end{figure*}

\begin{figure*}[ht!]
    \centering
    \includegraphics[width=0.7\textwidth]{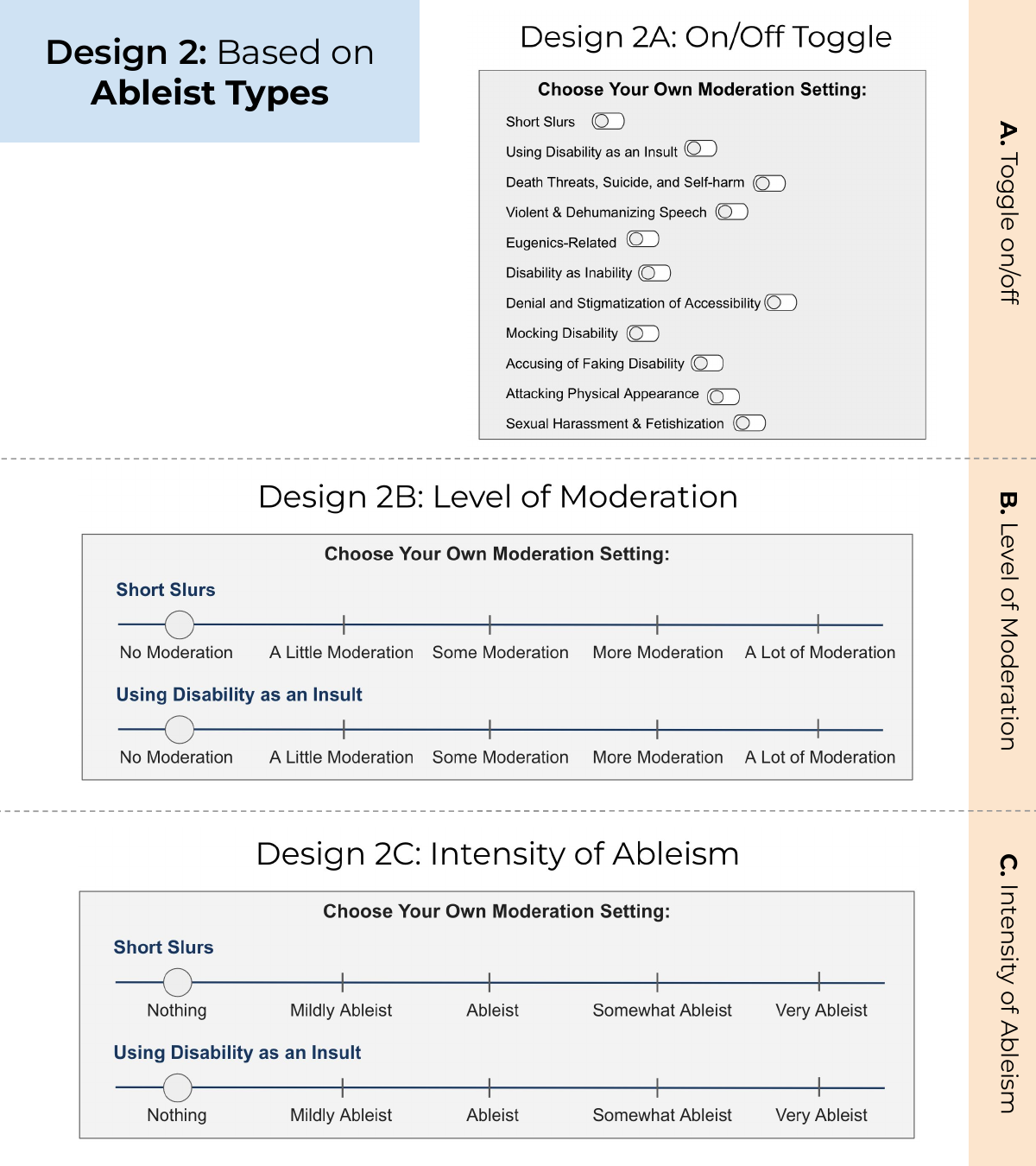}
    \caption{Diagram of Design Probe 2, configuring filters based on ableist types of hate. This includes 3 different designs: 2A (toggling each ableist type), 2B (slider for quantity of each ableist type), and 2C (slider for intensity of each ableist type). To reduce cognitive load, we asked participants to imagine design 2B and 2C to be applied to all the ableist types.}
    \label{fig:3}
    \Description{This diagram is titled “Design 2: Based on Ableist Types.” There are three designs shown. On the top, there is “Design 2A: On/Off Toggle” Underneath there is a grey box. In the box on the top, it reads "Choose Your Own Moderation Setting.” Underneath on the left hand side, there is a list (from top to bottom): “Short Slurs”, “Using Disability as an Insult”, “Death Threats”, “Suicide, and Self-harm”, “Violent & Dehumanizing Speech”, “Eugenics-Related”, “Disability as Inability”, “Denial and Stigmatization of Accessibility”, “Mocking Disability”, “Accusing of Faking Disability”, “Attacking Physical Appearance”, and “Sexual Harassment & Fetishization”. Next to each type there is a toggle on/off switch. In the middle, there is another design titled “Design 2B: Level of Moderation.” Underneath there is another grey box. Within the grey box it says Short Slurs. Underneath there is a slider that can be adjusted from (left to right): "No Moderation” to "A Little Moderation", "Some Moderation", "More Moderation", "A Lot of Moderation”. The slider is set to No Moderation. There is another slider with the title: “Using disability as an insult”. Underneath that, there is another slider which is the same as the slider described above (“No Moderation” to “A “ot of Moderation”). The slider is also set to “No Moderation”. At the bottom, there is the final design titled “Design 2C: Intensity of Ableism.” There is a grey box. Inside the grey box, on the top it reads: "Choose Your Own Moderation Setting." Underneath it says “Short Slurs.” Below that there is a slider (from left to right) that says: "Nothing", "Mildly Ableist", "Somewhat Ableist", "Ableist", and "Very Ableist." The slider is set to “Nothing”. There is another slider titled “Using disability as an insult”. Underneath that there is another slider which is the same as the slider described directly above (“Nothing” to “Very Ableist”). The slider is also set to “Nothing”.  On the right side there is a yellow vertical label categorizes the designs into three sections: A. Toggle on/off, B. Level of Moderation, and C. Intensity of Ableism. Each section aligns with the three different types of designs. }
\end{figure*}

\textbf{Filter Configuration (Design Probe 1 \& 2).} The first set of design probes featured varying filter interfaces for configuring user preferences. Design 1 (see Figure \ref{fig:2}) allows users to configure based on ableism, similar to current toxicity scales \cite{jhaver23personalizing}. Design 2 (see Figure \ref{fig:3}) allows users to configure based on types of ableist hate, using Heung et al.'s taxonomy of ableist hate and harassment \cite{heung24vulnerable}. Within Design 1 and Design 2, we presented varying control elements, similar to Jhaver et al.'s \cite{jhaver23personalizing} personalized moderation designs for toxicity. This included: A) toggle (on/off functionality), B) a slider on the proportion of moderation (percentage of ableist posts randomly removed), and C) a slider on the intensity of ableism. These design probes provoked preferences on ways to configure given levels of granularity and labor as well as participants' overall perceptions of using AI to filter ableist content.

\begin{figure*}[ht!]
    \centering
\includegraphics[width=1\textwidth]{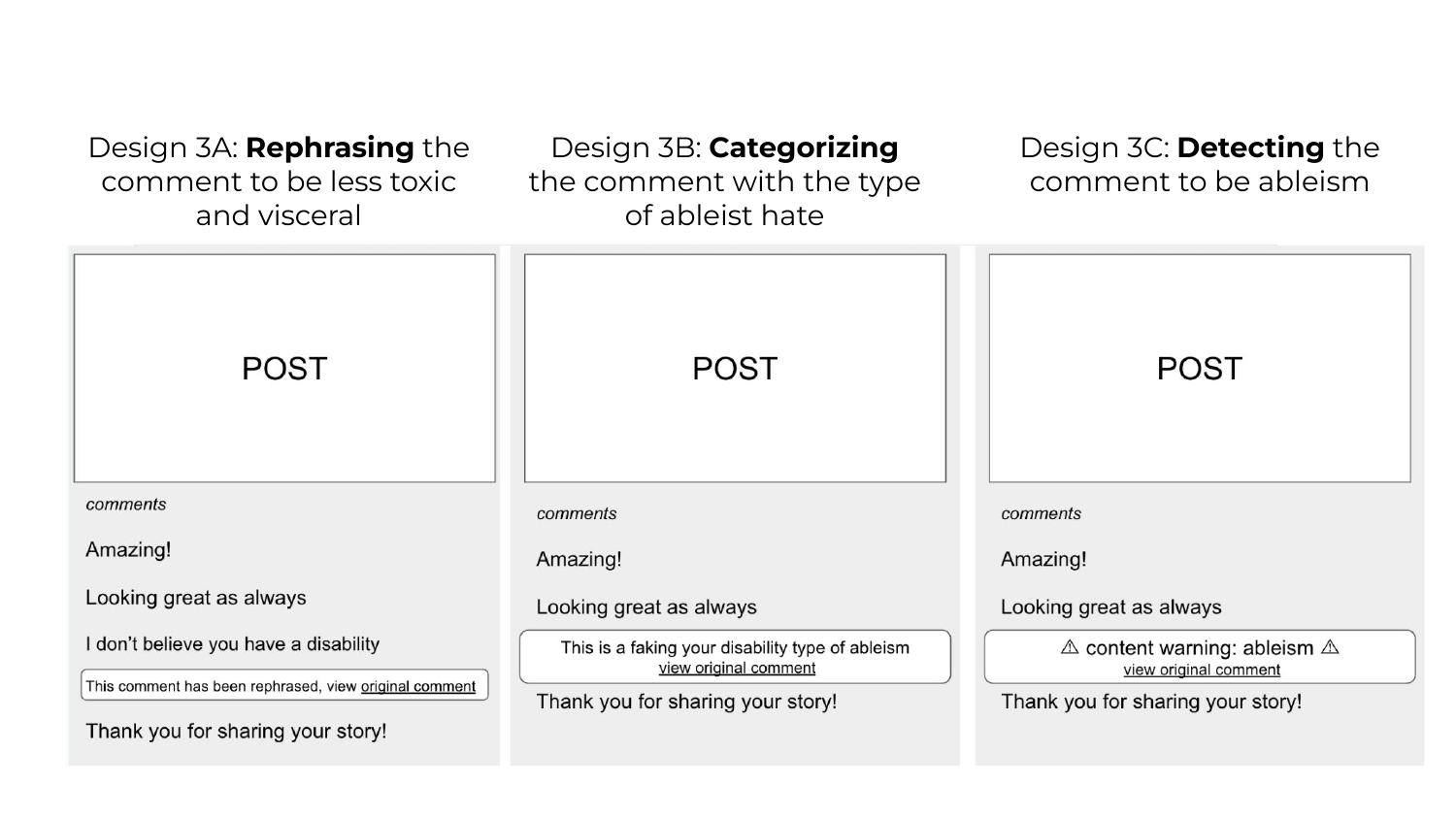}
    \caption{Diagram of Design Probe 3, customizing the presentation of the filtered hate. The original hateful comment contained slurs and an accusation that the user was faking their disability. We presented three designs to obscure or describe the hate, in contrast to current filters which completely remove moderated comments: 3A (AI rephrasing of the hate), 3B (content warning categorizing the type of ableist hate), and 3C (general content warning of "ableism"). For each design, users have an option to click and view the original comment.}
    \label{fig:4}
    \Description{This diagram shows three designs. The first (on the left) is titled “Design 3A: Rephrasing the comment to be less toxic and visceral”. Underneath there is a mock-up of a social media post with comments below. Underneath the post there’s a list of comments. Comments include "Amazing!" and "Looking great as always.” One comment says “I don’t believe you have a disability”, and right underneath there is an outlined bubble that says: "This comment has been rephrased, view original comment”. The original comment is underlined, to show that you can click into that link. The last comment says “Thank you for sharing your disability”. The next design (in the middle) is titled “Design 3B: Categorizing the comment with the type of ableist hate”. Below there is a similar mock-up of a post. The comments include "Amazing!" and "Looking great as always.” One comment has an overlay that says “This is your faking disability type of ableism”, and underneath it says “view original comment,” which is underlined. The last comment says “Thank you for sharing your disability”. The last design on the right is titled “Design 3C: Detecting the comment to be ableism”. Below there is a similar mock-up of a post. The comments include "Amazing!" and "Looking great as always.” One comment has an overlay that says “content warning:ableism”. There is a warning icon (triangle with an exclamation point inside). Underneath there is an option to “view the original comment,” which is underlined. The last comment says “Thank you for sharing your disability”.}
\end{figure*}

\textbf{Presentation of Hate (Design Probe 3).} The third set of design probes (see Figure \ref{fig:4}) explored personalizing the presentation of the hate, specifically augmenting the visibility of the hateful comment. Prior work has found social media users are hesitant to set restrictive filters due to their fear of missing out \cite{jhaver23personalizing}. Furthermore, disabled users, particularly creators and advocates, may want to read hate in order to make new advocacy content. We thus presented three design probes with varying levels of visibility to the original hateful comment. This included: A) rephrasing hate \textit{(highest level of visibility of the original comment)}, B) a content warning categorizing the type of ableist speech, and C) a content warning detecting ableism \textit{(lowest level of visibility of the original comment)}. The content warning design was inspired by prior work which found that trans users preferred customizable content warnings on social media \cite{haimson20trans}.

\subsection{Analysis}
Two of the authors analyzed the data by first preparing the transcriptions, listening to the audio recording to fix any issues with the automated transcriptions. We coded one interview and one focus group separately, then came together to discuss the codes. We repeated this process again with another interview and focus group recording. Through collaboratively discussing codes, we settled on a preliminary codebook, which included descriptive codes and codes tagging which design probe participants were responding to. We then split up the remaining transcripts, discussing new codes, consolidating codes with similar meanings, and collaboratively refining the codebook into categories (e.g., “concerns” and “wants”). We then conducted thematic analysis \cite{braun06using}, clustering similar codes together on a digital board and finding themes across categories of codes (e.g., "ableism is ambiguous" and "viewing hate is important for safety"). All authors participated in peer debriefing as interviews and focus groups were conducted and gave feedback on the codebook, clustering of codes, and themes. The iterative discussions throughout data collection and coding led to consensus.

\subsection{Positionality}
When conducting research on online harms, especially given the disproportional effect of hate and harassment on historically marginalized populations, it is essential to reflect as researchers. All authors have experience conducting research with and for the disability community and some members of the research team are disabled. We value amplifying the perceptions and experiences of disabled people within conversations on platform moderation and online safety. While personal moderation tools can be beneficial in mitigating harm caused by viewing ableist hate, such tools are not a solution to ableism, nor do they absolve a platform of their responsibility to remove harmful ableist content.
\section{Findings}
In this section, we share themes from the interviews (RQ1) and focus groups (RQ2). First, we present participants’ experiences addressing ableist hate and harassment online. We highlight their experiences using existing personal moderation tools, including blocking and word filters (Section 4.1). Then, we share participants’ perceptions and preferences on the design of an ableism-specific AI filter (Section 4.2 and 4.3). Throughout these sections, we interweave their perceptions on AI filters and their capability to identify ableist text. Lastly, we share how personal moderation has limitations in effectively reducing the harm of ableist hate (Section 4.4). Throughout our findings, we note which participants identified as content creators (C\#).

\subsection{Addressing Ableist Hate \& Harassment Online}

\subsubsection{Responding \& Educating}
Some participants responded to perpetrators of ableist hate by educating, especially if participants’ thought it was rooted in genuine ignorance, \textit{“a misunderstanding,”} (C18) or a result of someone being \textit{“uninformed”} (C9). However, participants were also wary of the risks of educating, which could escalate the hate further. C12 explained how publicly educating someone led to further harassment from others: 

\begin{quote}
\textit{"I was trying to be helpful by informing that it would be best to avoid that term [hearing impaired], because most people in the deaf community prefer... deaf or hard of hearing. And that person said, ‘okay thank you.’ But other people came in saying,’ that's a lie’, ‘that's not true’... [and] DMs saying, ‘oh, you're hearing impaired... Did I hurt your feelings? Lol.’”}
\end{quote}

A few participants responded to ableist hate with humor and satire. C18 described how she used her \textit{“vulnerability and sense of humor”} to effectively educate. C18 recalled an instance when someone said to her \textit{“why don’t you just stay at home?”}, and in response C18 created a post showing her accomplishments and \textit{“out having fun”} as a way to \textit{“reclaim it.”} C23 explained that she uses humor to showcase her resilience towards hate.

\begin{quote}
\textit{"I try to embarrass them and it's kind of funny... He's like, “I'm sorry no one wants your crippled p*ssy” and my response was, “that's not what your daddy said, so you need to stop before I give you another sibling and cut you out of the will.’... I just need people to know that I don't take this seriously.”}
\end{quote}

On the other hand, many participants explained that responding directly to harassers was not worth their energy and likely will not be productive. A few participants felt that they do not have the \textit{“bandwidth”} (C22) to be involved in sustained dialogue with harassers, which frequently led to \textit{“comment wars”} (C9) with no resolution. Furthermore, some participants refused to respond to harassers due to the social media algorithm giving hate more engagement, especially when \textit{“some accounts thrive on interacting with ableism”} (C7). C11 explained that responding could even benefit harassers because \textit{“it’s giving them engagement”} and \textit{“the fact that they've got a response means that they're probably more likely to... do it to other people who aren't as good as dealing with it."}

\subsubsection{Blocking}
All participants have blocked harassers before, noting it to be \textit{“the best option for your own health”} (C9). Blocking provided a form of relief to participants. P21 explained that by blocking \textit{“you’re addressing the issue... [the harasser] can't see me in any content or situation, and vice versa, and that's great.”} Participants also shared that blocking stopped harassers from \textit{“sending harassing content”} (C14) and helped participants \textit{“avoid future hateful comments”} (C5). 

Participants also shared ways blocking could prevent hate from other users beyond the harasser. For content creators, blocking harassers prevents them from \textit{“negatively influencing [the creators’] followers,”} (C5) and prevents hate from escalating. Another participant explained that blocking could help \textit{“control the reach of posts”} (P13). P13 would block someone if a harasser quote tweeted about him in order to stop the post from reaching their harassers’ followers, preventing potential hate from other users. 

On the other hand, some participants explained the downsides of not viewing the blocked person’s content, especially if the harasser was \textit{“talking about [them]”} and saying \textit{“things that are not true”} (C16). C5 explained that blocking meant they could no longer educate them on disability. C11 admittedly unblocked a harasser to inspire new advocacy content: \textit{“I have unblocked someone... to see what content I could create to almost combat it, which maybe isn't the healthiest thing... [but] I know that other people are seeing this, and I want there to be another side to the argument.”} Creators may choose to unblock or refrain from blocking harassers in order to counter ableist hate and engage in advocacy, despite the potential emotional or psychological toll it may take.

Participants also shared \textit{“loopholes”} (C14) harassers used to circumvent being blocked. Participants mentioned harassers could avoid account bans by creating new accounts and could \textit{“avoid the IP ban”} (C12) by using a VPN to modify their IP address. C14 expanded on harassers’ strategies of creating new accounts: \textit{"there's an account called ‘I hate wheelchairs’... They keep creating new accounts [and] they're on ‘Ihatewheelchairs4’... I don't understand how Instagram... allows that username to be made."}

While most participants felt blocking resolved hateful interactions, a few participants shared that blocking their harasser may escalate the hate from other social media users. This was especially worrisome for creators who had harassers with a large following. For example, C17 expressed a concern of blocked accounts screenshotting that they were blocked as a \textit{“flex,”} making it likely for \textit{“their followers to harass you”}.

Other participants expressed additional features that would enhance their experience with blocking. C9 and P15 requested a \textit{“block and remove interactions feature”} where they could block and remove all public interactions with that account, including posts the blocked account had liked or comments on posts. They suggested this feature would avoid being triggered by the blocked account's username, when looking through past posts. P21 desired a feature that would allow them to block someone without completely removing them from the Facebook group. They explained that the blocked individual may still have the right to be part of the group, but P21 wanted to block that account for her own "safety and sanity."

\subsubsection{Word Filters}
The majority of participants did not have experience using word filters, and of those who had used word filters most of them were creators. Some participants who used word filters inputted phrases to remove ableist hate, such as the r-word (C17, C12) or \textit{“faker”} (C14) to account for accusations that they are faking their disability. C14 also input her personal information into these systems to avoid being doxxed\footnote{being harassed by revealing someone’s private information without their consent}. She detailed how using word filters was laborious: \textit{"I'm constantly updating [my word filters] and adding different variations ofa... annoying ableist phrase.”}

A few participants used word filters for non-hate related moderation (C11, C16, P21), like spam. For example, C11 explained that when she made a post with \#chronic-illness, she received spamming comments invalidating her disability identity, saying \textit{"doctor so and so can heal you with his herbal blah.”} C11 then added "herbal" to her word filters; however, she was wary that genuine comments related to herbal tea could be filtered out.

Participants who used word filters noted flaws in the system, including filters hiding non-hateful comments while failing to block hateful comments. For example, C12 used a preset moderation filter on Twitch, and found it to filter out words he did not feel were offensive. While C12 noted the false positives, C16 entered keywords of a repeated comment she was getting from a pornographic account, but the word filters were not effective in stopping the account from commenting. C16 blocked some of those accounts but she \textit{“can’t block them all off [because] it’s a lot of work.”} 

While most participants shared experiences on mainstream social media platforms, C9 described her experiences of Tinder’s version of a word filter. She liked the design of first identifying potentially inappropriate words and then prompting the user if they want to block and report. 

\begin{quote}
    \textit{“If a guy... says ‘Can I see a nude photo?’ Tinder will ask you if the message makes you uncomfortable and if you say yes, it reports the message and blocks the dude... I like that cause it tells me right away, it's okay for you to report this... [and]} it gives you the ultimate say, ‘are you uncomfortable?’... I like that it gives people agency.”(C9)
\end{quote}

\subsection{Perceptions on AI Filters Identifying Ableism}
In response to Bleep’s interface, Design Probe 1 (configuration via ableism), and Design Probe 2 (configuration via ableist types), participants shared their perceptions and preferences on configuring AI filters to identify ableist text. Furthermore, participants revealed their concerns on AI's capabilities of correctly identifying ableism, shared implications of AI's inaccuracies, and brainstormed design suggestions to account for AI’s inaccuracies.  

\subsubsection{Preferred Configuration Design: Ableist Types}
Overall the majority of participants preferred configuring their filters via ableist types (Design Probe 2) because it was granular and understandable as to what the filter would be removing. Before viewing Design Probe 2, a few participants suggested a similar design that allowed them to filter out specific forms of ableist speech. P13 suggested a checkbox design that could allow him to input his preferences based on what he defines as most hurtful.

\begin{quote}
\textit{“Rather than a sliding scale, boxes that you could check for... would be more helpful. Because patronizing comments... get to me more than just straight up hate speech cause I can just... block... But the patronizing comments... [I have to] engage with you now, teach you something... they can take a lot more time and energy...They [the comment] might be not as ableist, but they might not be the lowest impact."} (P13)
\end{quote}

In response to Design Probe 2, participants shared how they would configure their filters, highlighting how some types of ableist speech are more emotionally draining than others. For instance, C9 prefers configuring by ableist types, because she can filter with a level of specificity that aligns with her personal triggers and preferences.

\begin{quote}
    \textit{"What matters... [is] the actual content of what's being said... If someone makes a patronizing comment to me, I'm just gonna make a joke back... but... I really really feel uncomfortable when I'm doing a nail tutorial and I have people saying sexually fetishizing things, because, A) my parents follow me… B) I just don't want to see it, because it makes me not want to post photos of myself... I sit there and think is that what everyone is thinking of me?... It's the intent of what is said that matters... it makes me feel unsafe}." (C9)
\end{quote}

For this very reason, the majority of participants expressed ableism (Design Probe 1) to be too broad of a concept, too vague, and expressed hesitancy to use the filter because they were unsure what was being filtered out. For example, after viewing 1C (intensity slider of ableism), C18 explained \textit{"I don't think that... [an intensity]} scale... would necessarily accomplish...instances [when] I do wanna see less of something." Conversely, a few participants were comfortable with the vagueness of “ableism” and preferred a straightforward method for filtering out ableist speech. For example, P2 appreciated Design Probe 1A, which allowed him to simply \textit{“toggle it on and [he’s] good to go.”}

\subsubsection{Preferred Control Elements}
Participants shared their varied preferences for control elements (A: toggle, B: slider of moderation percentage, C: slider of intensity) for each configuration design (Design 1: configuration based on ableism, Design 2: configuration based on ableist types).

\textbf{Toggle vs. Sliders.} Overall, participants preferred toggles over the sliders. Some participants liked design 1A, since it was  \textit{“easy to navigate”} (P4) and they did not want to view any sort of ableism. The majority of participants wanted more control over what types of ableist content they wished to not view, so they felt Design 1A was\textit{“too vague”} (C11). Therefore, they preferred using toggles to configure via ableist types (Design 2A) as that allowed them enough granular control to decide whether or not they wanted to view that type of ableist speech. C23 expanded on why she preferred an on/off mechanism: 

\begin{quote}
    \textit{“I won't have to battle within myself over the intensity of what I want to see... it's just all or nothing... [Do] I need to face reality?... What days can I accept this harassment and what days do I want to fight it?"} (C23)
\end{quote}

Similarly, C10 elaborated on why a toggle may be more fitting for certain types of ableist hate.

\begin{quote}
    \textit{“I hate when people say I'm faking a disability or... that I don't 'look disabled.' I would not want: slightly, mildly, a little, [or] a lot. I would want none at all... I feel like the sliders don't make a lot of sense, because... why would I want to let some of this still come through when it's the most upsetting thing to me?”}
\end{quote}

Several participants also noted that a toggle was more accessible for those who use \textit{“screen readers or who might have hand dexterity issues”} (C11).

\textbf{Moderation Slider (Design 1B/2B) vs. Intensity Slider (Design 1C/2C).} While some found "percentage of moderation" to be an easily understandable and objective measurement, a majority of participants felt that it did not make sense for mitigating harm. P21 explained: \textit{“it’s just randomly selecting 25\% or 50\%... that doesn’t seem like it’s... moderating.”}

Participants generally preferred the intensity slider over moderation, but several participants also expressed flaws with the intensity slider. In the focus group, C7 pointed out that one \textit{“cannot quantize disability, accessibility, or ableism.”} Using humor, the focus group attendees explained how quantifying ableism minimized and trivialized the harm done. C7 said: \textit{“mildly ableist or somewhat ableist sounds weird to me. It sounds like a joke...‘hey, someone punched me in the face.’ And they were like, ‘yeah but did they mildly punch you, or did they really punch you?"} 

Other participants also explained that \textit{“degree of ableism”} (C18), referring to Design Probe 1C, does not allow them to pick the content they personally find hateful. This is why many participants preferred ableist types (Design Probe 2). 

\subsubsection{What is Ableist is Contextual, Equivocal, and Contested}
Participants were skeptical if AI filters could accurately identify ableism due to 1) the subjective nature of what is ableist, 2) the nuance and context of when something is ableist, and 3) perceived biases and capabilities of AI.

Participants questioned if there is a universal understanding of what is ableism or what is ableist. This was a common concern, especially when configuring filters by ableism (Design Probe 1). Participants were unsure what types of content would be filtered out. C9 explained that ableism is \textit{“a broad category... not everything is the same level of problematic to each individual person.”} Participants shared examples of how, even within the disability community, there are disagreements on what is ableist. Participants referred to diverse preferences regarding self-identifying language (e.g., people with special needs, disabled people vs. people with disabilities, hearing impaired vs. deaf and hard of hearing). C7 and C16 explained that if AI were to block words related to one’s identity, this could cause fear of being \textit{“penalized”} for your own identity: 

\begin{quote}
    \textit{“I have quite a few friends who identify as hearing impaired and impaired is a very disgusting word in the disability space. But if she identifies that way, they are allowed to... I worry that those types of voices will be moderated out."} (C16)
\end{quote}

Since ableist language is not clear-cut, C11 explained that:\textit{“language that might be considered offensive.... also might exist for a good reason... I think that's just important to note when you're training AI, [ableist language is] just not explicit or definitive."}

Participants also viewed ableism as intertwined, but not identical to body shaming. Intel’s Bleep interface combines ableism and body shaming into one category of hate. C9 expanded on why this design is problematic.

\begin{quote}
    \textit{“As someone who is both fat and disabled, [ableism and body shaming] are different issues... not all disabilities are physical or has to do with the body... It feels like a very narrow definition of disability when you also loop it with body shaming, because to me it looks like they're only looking at visible disabilities."}
\end{quote}

\subsubsection{Skepticism and Concerns with AI Accuracy} 
Since ableism is subjective and nuanced, many participants were skeptical if AI could accurately identify ableist comments. Participants were concerned that the filters would take keywords out of context, leading to false positives (AI wrongfully filtering out comments about disability). C19 explained that she does not \textit{“trust AI... to distinguish what is ableist and what is disabled or disability adjacent"}(C19). This included filtering out ableist terms used within historical, medical, educational context that are not intended to be hateful towards an individual, or if a user was recounting a hateful experience and directly quoted hate they had received. The potential of false positives caused participants to fear missing out on constructive conversations and opportunities to connect with the disability community. For example, C9 expressed uncertainty of how well AI can understand context:

\begin{quote}
    \textit{“Would you not be able to read historical things because they use words differently? If you're talking about a... time when... the [r-word] was acceptable... I feel like without there being a contextual component you're gonna miss, so many conversations that may not be what this thing [AI] thinks it is, but... I don't understand AI that well."}
\end{quote}

Similarly, participants were also concerned with AI filtering comments with disability-related reclaimed words (e.g., cripple, gimp). Participants agreed that what differentiates a slur from a reclaimed word was dependent on who said it and the intent. P13 questioned how AI filters could account for this context:

\begin{quote}
    \textit{"There's the tension within in-group and out-group language... not everybody has earned the right to use certain words. But within a group... there's safety and familiarity... One word that I would be really upset to hear from somebody who wasn't also disabled would be cripple. But that's a word I've definitely used... And it's a word that ... has scholarly usage... So how is the intent... accounted for?"} 
\end{quote}

A couple of participants anticipated the AI to be inaccurate due to ableist biases, assuming that AI models may not be trained on the lived experiences of disabled people. P13 anticipated the AI to over-filter content about disability and sex, because of stereotypes of disabled people being desexualized. Furthermore, some participants doubted AI’s accuracy due to negative experiences with online moderation. Creators were especially wary given their past, negative experiences with moderation.

\begin{quote}
    \textit{"When I try to post content... it'll be reported as hate speech because I'm addressing someone as an 'able-bodied Savior or a white Savior'... These AI filters end up blocking a lot of disabled creators from sharing very important information... some of us feel like we've already been burned by moderation."} (C23)
\end{quote}

Over time, this may \textit{“inadvertently chill or silence the speech of pro-affirming disability language like conversations about disability, or about diversity, or about bodies in general" (C18).} 

\subsubsection{Designing for Inaccuracies}
Implications of AI inaccurately removing disability-related content included missing out on constructive conversations and infringing on disability activism and mobilization. Participants viewed filters as a possible threat toward the community they built online. For example, C23 explained: 

\begin{quote}
    \textit{“I'm afraid that I'll miss things, and it'll harm relationships that I've built...disabled people have found kinship on social media. Because for a lot of us we're the only disabled person we've ever known. We live in communities that can be isolating. So I don't want to ever miss [out] because of a filter that I've set."}
\end{quote}

Participants proposed additional features to alleviate these concerns. Several participants wanted a feature to select certain accounts to be exempt from the filter setting. This way, participants would not miss out on their disabled friends’ posts, knowing that if they did use an ableist term it would not be triggering or considered hateful. C7 explained: \textit{“let's say C11 and I are interacting on Twitter, and I know that she uses the word cripple to self-identify. I would put her on ‘no moderation’ or ‘a little moderation’ if I know... the word cripple is going to be triggered by [the filter].”} 

Since participants were unsure how AI is defining ableism, they wanted oversight to how it was filtering content. This included viewing what the filter is removing to \textit{“determine, is this worth it? Is it working?”} through a \textit{“test button”} (P21). Some wanted an \textit{“undo button”} (P3, C22) to override any mistakes the AI makes, such as filtering out a friend’s comment that they would have otherwise responded to. 

Some participants wanted to train the AI to align with their preferences by tagging content they found as ableist.Participants also explained that it was important to communicate to the algorithm not only what is ableist vs. what is not, but also why. 

\begin{quote}
    \textit{“If AI learns from many actions, then the more information we can give it, probably the better. So if one is accepted or allowed through [the filters], the why or the justification could be: ‘this is a person in this community with this identity has chosen to reclaim this language.’ And one that gets rejected might be ‘this is a person… harassing someone’... I would probably type 2 or 3 sentences... [but] I don't know if the AI would understand that."} (C20)
\end{quote}

C18 added that teaching the AI allows each user to \textit{“define for themselves what ableism means and looks like.”}

\subsection{Varied Tolerances on Viewing Ableist Hate}
In response to Design Probe 3, participants shared their own individual preferences of if (Section 4.3.1) and how to view ableist hate (Section 4.3.2 \& Section 4.3.3), acknowledging that other disabled users have \textit{“different levels of tolerance”} of viewing ableism (P13).

\subsubsection{To View or Not to View Hate}
Participants shared their philosophies of whether or not to view hate. A couple of participants prioritized protecting themselves from viewing hate online, wanting the ability to turn off all hate. For example, P3 explained that \textit{“we are responsible for our own safety... I would rather just toggle the entire thing out.”} 

Some participants explained that, while in theory they wouldn’t want to view hate, in practice they would be too curious to not. Although participants appreciated having the option to view the original comment in Design Probe 3, a majority of participants acknowledged that having an option to \textit{“view original comment”} was too tempting to not click. C17 explained:

\begin{quote}
    \textit{"I'm just always so curious... I would probably say 80\% of the time. It's just gonna bug the hell out of me if I don't know what is being said, because… I need to have that control. But then there's gonna be that other 20\% [where] I just don't feel like it today.”}
\end{quote}

Due to inherent curiosity, a couple of participants wanted the \textit{“view original comment”} button to be removed. 

Several creators stressed the importance of knowing what is being said about them in order to \textit{“control the narrative”} and felt a strong responsibility to moderate their own page (C9). C9 explained that if other users are spreading rumors and talking about him on his own channel, then \textit{"ignorance is not bliss."}

While hate is harmful, a few participants emphasized the importance of viewing hate for their own physical safety. C23 was concerned about hiding death threats: \textit{“if it’s a serious threat, someone should be notified, whether it’s [the] police or anyone."} P13 highlighted the trade-off of protecting themselves from hate and being aware of potential dangers. 

\begin{quote}
    \textit{“As much as I don't like seeing hate speech, it is helpful to know the conversations that are happening right, what kind of is the Zeitgeist... and to just automatically screen that out also doesn't really contribute always to a sense of safety. Because it means you actually can't be aware of potential dangers"} (P13)
\end{quote}

These participants highlighted the need for personal moderation tools to balance protection from hate with awareness of serious risks, such as death threats. 

\subsubsection{Rephrasing Feels Fake, Patronizing, and not Helpful.} The majority of participants' initial reactions towards rephrasing was negative, finding AI rephrasing hate as uncomfortable and dishonest. C9 explained how the rephrased version of hate \textit{“feels fake… I'd rather know what somebody just said so then I can report it.”} (C9) Furthermore, several participants discussed how rephrasing hate felt \textit{“patronizing [and] infantilizing”} as if participants needed protection from AI to give them the \textit{“nice version”} of hate (C7). A few participants were skeptical of AI’s ability to generate an \textit{“authentic translation”} and rephrasing of the original comment (C19). 

Participants also added that it’s not worth rephrasing hate if everyone else online can still view the original hate. P8 explained that \textit{“it [3A] doesn't really make sense to see a softer version of [an] insult and everyone knows in the comment section it’s saying something else... I don't feel it’s necessary. If it’s hateful, let me see it that way."}  

Many participants explained that rephrasing hate was largely not effective, since it does not change the intent of the harasser. C23 called rephrasing ableist hate as \textit{“diet ableism, it's diluted [and] it'll hurt less”} but \textit{“it's not sparing anybody's feelings.”} 

Several participants suggested the rephrasing function should instead be used to educate harassers. C18 explained that while rephrasing helps with \textit{“softening people’s hatred,”} she feels it reduces accountability by \textit{“letting [harassers] off the hook for their hateful language.”} Additionally, P21 clarified why she rather have the system to be used to educate harassers:

\begin{quote}
    \textit{“[3A] seems like a strange way of trying to protect our feelings, which being disabled, I don't want you to try and protect my feelings like I will take actions to do that... if everybody else can still see the damaging words, then… What is the point of this? ”}
\end{quote}

\subsubsection{Content Warnings Are Informative.} The majority of participants preferred content warnings (Design Probe 3B \& 3C) before viewing ableist hate, though their preferences on the amount of detail about the hate varied. Some participants preferred 3B (categorizing hate) because it’s \textit{“helpful… to get to know the information [about the hate] but not feel attacked"} (C5).

Categorizing the hate empowered the user to make an informed decision of whether or not to view it. Since the comment is not removed, the user has agency to view it at any time. For example, C16 said \textit{"I like that it gives me the option to look at it or not, and I can decide based on my mood at the moment… so I can look at it later, when I can handle it."} C7 emphasized the importance of designing for autonomythat was missing when AI rephrased hate for them.

\begin{quote}
    \textit{"Many disabled people have that autonomy taken [away] from them, or are told they don't know how to make decisions or patronized... [a content warning]is more like a trigger warning where you are given the autonomy, whether you want to read it... That's not taken away from you as it was in [Design Probe 3A]."}
\end{quote}

A few participants added how categorizing the hate improved the explainability of the filter system. C10 described how the category is not only \textit{“giving an explanation as to what kind of comment was said”} but also explaining \textit{“why the AI detected and flagged the comment.”} C11 elaborated that the explainability built more trust with the AI and reduced the likelihood of viewing the original comment. 

\begin{quote}
    \textit{"If it's just “ableism” (referring to Design Probe 3C) again for that morbid curiosity thing, you might want to read it, or you might want to check whether it actually is ableist or not, whereas if it's more specific... there's a little bit more trust that the AI knows what it's doing if it can detect the category, whereas if it's just ableism, the trust is less.”}
\end{quote}

On the other hand, some participants preferred a general content warning (Design Probe 3C), because it is enough information to make a choice of whether to view the content. A few participants added that categorizing ableist hate (Design Probe 3B) may be too triggering. For example, C16 said: \textit{"I have mixed feelings about announcing the type of ableism, because… on a bad day I don't even wanna know that it was ‘your faking disability’ type of ableism."} With a similar sentiment, C5 and C12 preferred Design Probe 3C since it was a generic warning with less risk of being emotionally triggered.

\subsection{Limitations of Personal Moderation}
While some participants shared how filters have the potential to reduce harm of viewing ableist content, other participants wanted filter settings to be applied to everyone’s view, not just their own view. Having the hate viewable for others may cause harm to others or escalate hate. Participants felt that inadequate moderation for ableism seemed unfair since other types of discrimination were being removed by the platform.

\begin{quote}
    \textit{“Does it [the filters] protect me more? Yes. But if people can engage with it, comment on it, or view it potentially. I don't like that... Other hate speech that is more widely recognized is automatically deleted most times on other posts so why is Ableism an exception.”} (C14)
\end{quote}

Participants added how the various presentations of the hate (Design Probe 3) could be educational and therefore wanted all social media users to view it. More specifically, participants shared that the filters based on ableist types and the detailed content warnings could spread awareness of the different types of ableism. For example, P21 liked the ableism content warning and wanted everyone to be able to view the content warning as a way to educate others that ableism exists. P21 explained how she imagined a teaching tool to be designed using all of the different versions of Design Probe 3.

\begin{quote}
    \textit{"[Design Probe 3] should lead with... ‘content warning: ableism’, and then underneath it say[s] ‘this is the faking your disability type of ableism.’ And then under that ‘this comment has been rephrased to say, I don't believe you have a disability.’ And then underneath that it would say, ‘view original comment.’ So then you're teaching everyone...[other users are] learning... what is ableism,... the type of ableism, and... how to rephrase something.”}
\end{quote}

C18 added that the original poster should be notified if their content is filtered out by another user and \textit{“tagged as this type of ableism,”} so the harasser can be educated on why it’s ableist. 

Other participants acknowledged the limitations of filters: it does not hold users accountable for perpetuating ableist speech. Therefore, a few participants recommended the filter system to trigger repercussions like a suspension. P13 explained that \textit{“personalized moderation isn't really a solution to unsafe communities and unsafe spaces… [There] needs to be a community responsibility as well.”} Participants felt that \textit{“community responsibility”} needed to come from the platform moderation itself by seriously addressing reports of ableist speech. C23 attributed lack of moderation of ableist hate due to lack of knowledge on ableism:

\begin{quote}
    \textit{“Harassing disabled people is the one thing you can get away with pretty easily on social media... Attacking someone's disability isn't seen as a problem because most of the moderators aren't disabled. They're never going to challenge comments like, ‘you shouldn't be proud to be disabled’... [because moderators] themselves have those same beliefs."}
\end{quote}

\section{Discussion}
While personal moderation enables users to control for various types of content, ranging from content that is harmful to content that is uninteresting, our study examines one specific type of content users may prefer to avoid: ableist hate. Given this context, we present design recommendations for an ableism-specific AI filter to support safety, harm reduction, and agency. We also make recommendations related to usability, explainability, and trustworthiness of the system, as it may impact whether or not users adopt AI filters. Lastly, we share our study’s limitations and directions for future work. 

\subsection{Design Recommendations for Personalized Moderation}
Our design probes elicited participants' values when using a personal content moderation tool during their experiences with ableist hate online. We discuss design recommendations that support values participants' shared (refer to Table \ref{tab:designrecs} for a summary).

\begin{table*}[h] 
\small
\caption{Summary of Design Recommendations for Personal Content Moderation.}
\label{tab:designrecs}{%
\begin{tabular}{@{}ll@{}}
\toprule
\textbf{Value \update{\footnotemark}} & \textbf{Design Recommendations}                                                            \\ \midrule
Promoting Safety                   & Design notifications pertaining to threats to one’s physical safety of the user rather than removing them from view \\ \midrule
Enhancing Explainability \& Usability & Use ableist types for configuring personal content moderation settings                     \\ \midrule
Supporting Agency \& Reducing Harm & Embed content warnings as an option to support users’ decision making in deciding whether or not to view hate       \\ \midrule
Building Trust                        & Implement ways to oversee filtering, undo-decisions, and support AI learning from the user \\ \bottomrule

\end{tabular}%
}
\end{table*}

\update{\footnotetext{Values listed are equally important and represent a design space rather than an ordered list.}}

\subsubsection{Threat Notifications Promote Safety}
Since personal moderation tools are designed to protect users online, it's essential to address situations where they may unintentionally create other safety risks. As noted by our participants, while personal moderation might enhance psychological comfort by removing ableist speech from view, it can compromise safety by reducing a user's awareness of potential dangers.For example, moderation tools may completely filter out hate related to physical safety, such as death threats. Tune, Google’s content moderation tool, has a disclaimer of this limitation: “Tune isn’t meant to be a solution for direct targets of harassment (for whom seeing direct threats can be vital for their safety)” \cite{tune24}. Removal of threats to one’s safety may lead to ignorance of unsafe physical spaces and events and lead to in-person harms \cite{scheuerman18safe}. This is increasingly relevant as social media has become a central information hub for news and events \cite {pew23social}. Additionally, eliminating ableist hate might obscure the cultural climate or "zeitgeist." This may lead to disabled individuals being unaware of potential risks of harassment related to disclosing their disability  \cite{eagle23you, heung22nothing, heung24vulnerable} and emotional harms to their self-esteem and possible internalization of disability stereotypes \cite{ringland19understanding}.

Removing ableist speech may also reduce the agency of disabled people. This study and prior work has shown how both disabled and non-disabled social media users alike use blocking, reporting, and responding strategies to address hate \cite{samermit23millions, eagle23you,thomas22its, heung24vulnerable, heung22nothing, musgrave22experiences, harris23honestly}. Our participants were hesitant to use filters, given that it could prevent them from using blocking, reporting, and responding to reduce anxiety and prevent hate from escalating. For example, blocking prevented repeated harassers posting ableist comments. Content creators or high-profile users wished to control narratives about themselves, as leaving hate could cause reputational damage \cite{thomas22its}, lead to more engagement and hate from other users due to the platform algorithm \cite{bertaglia24the}, and subsequently may escalate into in-person harms \cite{maarouf24the}. 

Not accounting for the above use cases may lead to increased safety risks towards users from historically marginalized communities who are at-risk of identity-based hate. We recommend for personal content moderation systems to consider not removing threats to one’s physical safety, but instead design notifications that inform users of safety concerns without burdening users who are targets to hate \cite{thomas21sok}. Researchers should explore AI’s role in effectively implementing safeguards for users’ safety when using personal content moderation tools. 

\subsubsection{Ableist Types Enhance Explainability \& Usability}
Current personal content moderation tools often use broad categories like "toxicity" or “sensitivity,” which may lack clarity as to what exactly the system is filtering. Prior research \cite{jhaver23personalizing} highlights the need for more explicit definitions of what constitutes toxic and sensitive context. However, our findings suggest that definitions alone may not effectively communicate what is being filtered. The definition of "what is ableist" is hotly debated within the disability community, complicating how a system \textit{should} define ableism explicitly and how a system should categorize the varying degrees of ableism (e.g., mildly ableist vs. very ableist). Intensity of ableism (design C) may trivialize ableist speech and may invalidate individual perceptions of how ableist a comment is to them. Ableist types can be an alternative to communicate explicit forms of ableist speech without offending or invalidating the user, especially since our findings imply that how people experience ableism is subjective.

Detecting toxic language, either as a binary value (i.e., is it toxic or not?) or by measuring the intensity of toxicity, are common measurements and standard practice in machine learning-based moderation systems (e.g., Perspective API \cite{perspectiveAPI}). However, prior work has highlighted biases in toxicity detection systems \cite{sap19the, oliva21fighting,garg23handling,goyal22is}, including ableist biases \cite{Hassan21,venkit23automated}. Given these existing concerns, we argue that personal moderation requires a shift from classifying hateful text based on toxicity levels to classifying it based on the type of hate found within the text (in our particular context, the type of ableist hate). Participants understood ableist speech as types, not as a numerical value. Enabling disabled people to choose ableist types may be more explainable and usable in configuring what kinds of ableist speech they wish to filter out. For example, some participants found patronizing comments more exhausting than outright ableist slurs and preferred not to see patronizing comments. If a system only affords the user to select severity of ableism, it may not account for personal preferences the user wishes to view and not view. Furthermore, designing the interface with these types may be more intuitive than using a slider that adjusts the intensity of ableism. Allowing users to configure filters based on ableist types aligns better with their preferences for filtering specific content they personally find to be more ableist or hurtful.  

\subsubsection{Content Warnings Support Agency \& Reduce Harm}
Similar to prior work on the usage of content warnings on social media \cite{haimson20trans}, our findings suggest users may benefit from content warnings that support informed decision making and control. For example, users could decide to view ableist hate if they were in the "mood," did not find the type of hate particularly triggering, and/or if they wanted to respond to the hate. Some appreciated the explainability of content warnings for categorizing the type of hate as this fostered greater trust with the AI system accurately identifying ableist hate. Participants noted that this clarification improved transparency, a highly valued characteristic among social media users regarding moderation tools \cite{jhaver23personalizing, ma23defaulting}. Additional transparency also alleviated concerns of missing out, giving more assurance that the AI was filtering ableist hate, not disability-related content.

Although content warnings were perceived as beneficial, it is important to consider the potential side effects of content warnings. While content warnings “allow those who are sensitive to these subjects to prepare themselves for reading about them, and better manage their reactions” \cite{manne25why}, prior work has detailed how content warnings can backfire. Content warnings may not be helpful for those experiencing trauma or re-traumatization \cite{scott23trauma}. Content warnings may cause a “forbidden fruit effect,” \cite{bridgland24meta} making the content seem more attractive. Participants anticipated this effect by explaining they would be “too curious” to not “view original comment” in Design Probe 3. Future work is needed to evaluate the varying designs and effects of content warnings on social media, especially for identity-based harmful content.

\subsubsection{Oversight \& Reversibility Features Build Trust}
Due to prior negative experiences with moderation, participants expressed skepticism and hesitation towards using AI filters. While previous research indicates that users are wary of AI filters over-moderating and fear of missing out on content \cite{jhaver23personalizing}, our participants raised additional concerns specific to the lack of widespread understanding of ableism. Furthermore, ableism is often poorly moderated; prior work has noted instances of wrongful removal of disability-related content \cite{heung24vulnerable, lyu24because} and instances of ableist hate not being addressed by platform moderation \cite{heung24vulnerable}. This aligns with research showing that existing mistrust in a domain extends to AI-based solutions (i.e., mistrust in moderation extends to mistrust in AI-based moderation) \cite{lee21who}. This skepticism may be applicable to other historically marginalized groups who have similarly felt unsupported by platform moderation, such as black people \cite{harris23honestly,musgrave22experiences,han23hate} and LGBTQ people \cite{oliva21fighting,han23hate,moolenijzer23they}. Consequently, those with negative experiences and distrust in current moderation systems may also view AI moderation tools with similar suspicion. This is particularly relevant for active social media users, such as creators, who are more likely to encounter negative moderation experiences.

Participants perceived filtering of identity-based harmful content by AI to be risky, especially due to concerns about AI over-moderating content highly relevant to their identity and daily life. Prior work suggests that mistakes by AI filters may be detrimental to disability advocacy, community building, and information gathering \cite{heung24vulnerable, rauchberg22shadowbanned,kaur24challenges,borgos-rodriguez21understanding}. Because of these concerns, our findings suggest that users may be reluctant to adopt AI filters without robust safeguards and fail-safes. Therefore, we recommend implementing features like the ability to undo and correct filtering errors so the underlying model can learn from these corrections. Additionally, to give greater oversight and control over \textit{who} the AI filters out, we also recommend implementing allowlists \cite{dunn23avoid}, a feature that allows users to select trusted accounts exempt from filtering. By adding such features, users may feel more confident in using AI filters. 

\subsection{Limitations \& Future Work}
Since we focused on showing design probes that were meant to provoke participants’ wants \cite{wallace13making}, we did not build an interactive prototype. This trade-off was intentional, as an interactive prototype may bias participants to only share what they think is feasible \cite{hakim00effective, rudd96low}. The technical feasibility of an ableism-specific AI filter may not be far off. For example, generative AI models like ChatGPT show promise of moderating user-generated  content, as researchers begin to evaluate AI’s effectiveness in identifying hateful content \cite{li24chatgpt}. Future research should build and evaluate such a tool, which may provide additional insights on user’s behaviors over time, usability considerations, and AI’s accuracy in identifying ableist speech. Furthermore, future work should investigate how the usage of personalized moderation varies across different user types (e.g., content creators vs. casual users) and among users with varying identities (e.g., disability, race, sexuality). As these tools aim to encourage safe participation online, it is important that they are accessible and do not contribute to the labor disabled users exert to be on social media \cite{lyu24because,heung24vulnerable, borgos-rodriguez21understanding,simpson23hey,mok23experiences,rong22it}. Furthermore, future work should investigate AI-based personal moderation for other forms of identity-hate and other manifestations of hate like image- and video-based hate. 

While AI-based personal moderation can contribute to a fairer and less harmful online environment for disabled users, it is not a substitute for structural changes needed to achieve justice for those affected by ableism \cite{bennett20what}. Future research should consider the role of AI-based platform moderation to combat structural ableism, such as designing for disability awareness and education. For example, participants wanted rephrasing to be a platform intervention, nudging perpetrators to reconsider posting ableist hate and learn more about ableism. However, it is critical to ensure that proactive nudges do not enable harassers in alternate ways; for example, actors can be even more toxic if subverting moderation systems becomes gamified \cite{warner24critical}. Future work should consider incentivizing positive and prosocial user behavior to prevent ableist hate in the first place, while protecting against the potential adverse effects of proactive nudges. 
\section{Conclusion}
This paper investigates how AI-based personalized moderation can safeguard disabled users from viewing ableist hate on social media. We created design probes to elicit users' preferences for an ableism-specific filter, including ways to filter ableist text (e.g., based on types of ableist hate) and ways to customize the presentation of hate (e.g., AI rephrasing hate or content warnings). We share design recommendations related to supporting users' safety (e.g., being notified of personal threats), improving usability (e.g., filtering based on ableist types), reducing the harm (e.g., content warnings) and building trust (e.g., undoing filter decisions). Lastly, we further conversations on personal moderation to address identity-based harms, amplifying the perspectives of disabled people using personal moderation tools for ableist hate. 

\begin{acks}
    We thank our participants for their thoughtful feedback during the focus groups and our anonymous reviewers for helping improve the paper. This work was supported in part by Google, President’s Council of Cornell Women (PCCW) and the National Science Foundation.
\end{acks}

\bibliographystyle{ACM-Reference-Format}
\bibliography{bib}


\begin{thebibliography}{96}


\ifx \showCODEN    \undefined \def \showCODEN     #1{\unskip}     \fi
\ifx \showDOI      \undefined \def \showDOI       #1{#1}\fi
\ifx \showISBNx    \undefined \def \showISBNx     #1{\unskip}     \fi
\ifx \showISBNxiii \undefined \def \showISBNxiii  #1{\unskip}     \fi
\ifx \showISSN     \undefined \def \showISSN      #1{\unskip}     \fi
\ifx \showLCCN     \undefined \def \showLCCN      #1{\unskip}     \fi
\ifx \shownote     \undefined \def \shownote      #1{#1}          \fi
\ifx \showarticletitle \undefined \def \showarticletitle #1{#1}   \fi
\ifx \showURL      \undefined \def \showURL       {\relax}        \fi
\providecommand\bibfield[2]{#2}
\providecommand\bibinfo[2]{#2}
\providecommand\natexlab[1]{#1}
\providecommand\showeprint[2][]{arXiv:#2}

\bibitem[ins(2021)]%
        {instagram21sensitive}
 \bibinfo{year}{2021}\natexlab{}.
\newblock  (\bibinfo{year}{2021}).
\newblock
\urldef\tempurl%
\url{https://about.instagram.com/blog/announcements/introducing-sensitive-content-control}
\showURL{%
\tempurl}


\bibitem[per(2021)]%
        {perspectiveAPI}
 \bibinfo{year}{2021}\natexlab{}.
\newblock \bibinfo{title}{Using machine learning to reduce toxicity online}.
\newblock \bibinfo{howpublished}{\url{https://perspectiveapi.com/}}.
\newblock
\newblock
\shownote{Accessed: September 10, 2024}.


\bibitem[aut(2023)]%
        {automodreddit}
 \bibinfo{year}{2023}\natexlab{}.
\newblock \bibinfo{title}{r/AutoModerator}.
\newblock \bibinfo{howpublished}{\url{https://www.reddit.com/r/AutoModerator/wiki/can_do/}}.
\newblock
\newblock
\shownote{Accessed: September 10th, 2024}.


\bibitem[pew(2023)]%
        {pew23social}
 \bibinfo{year}{2023}\natexlab{}.
\newblock \bibinfo{title}{Social Media and News Fact Sheet}.
\newblock
\newblock
\urldef\tempurl%
\url{https://www.pewresearch.org/journalism/fact-sheet/social-media-and-news-fact-sheet/}
\showURL{%
\tempurl}
\newblock
\shownote{[Accessed 11-09-2024]}.


\bibitem[bod(2024)]%
        {bodyguard24}
 \bibinfo{year}{2024}\natexlab{}.
\newblock \bibinfo{title}{Bodyguard}.
\newblock
\newblock
\urldef\tempurl%
\url{https://www.bodyguard.ai/en}
\showURL{%
\tempurl}


\bibitem[fac(2024a)]%
        {facebook24how}
 \bibinfo{year}{2024}\natexlab{a}.
\newblock \bibinfo{title}{How Do I Report Something I See on Facebook?}
\newblock \bibinfo{howpublished}{\url{https://www.facebook.com/help/814083248683500}}.
\newblock
\newblock
\shownote{Accessed: September 10, 2024}.


\bibitem[x24(2024)]%
        {x24blocking}
 \bibinfo{year}{2024}\natexlab{}.
\newblock \bibinfo{title}{How to block accounts on X}.
\newblock \bibinfo{howpublished}{\url{https://help.x.com/en/using-x/blocking-and-unblocking-accounts}}.
\newblock
\newblock
\shownote{Accessed: September 10, 2024}.


\bibitem[twi(2024)]%
        {twitchautomod}
 \bibinfo{year}{2024}\natexlab{}.
\newblock \bibinfo{title}{How to Use AutoMod}.
\newblock
\newblock
\urldef\tempurl%
\url{https://help.twitch.tv/s/article/how-to-use-automod?language=en_US}
\showURL{%
\tempurl}
\newblock
\shownote{[Accessed 11-09-2024]}.


\bibitem[tun(2024)]%
        {tune24}
 \bibinfo{year}{2024}\natexlab{}.
\newblock \bibinfo{title}{Tune (experimental)}.
\newblock
\newblock
\urldef\tempurl%
\url{https://chromewebstore.google.com/detail/tune-experimental/gdfknffdmmjakmlikbpdngpcpbbfhbnp?pli=1}
\showURL{%
\tempurl}


\bibitem[fac(2024b)]%
        {facebook24what}
 \bibinfo{year}{2024}\natexlab{b}.
\newblock \bibinfo{title}{What happens when you block someone on Instagram}.
\newblock \bibinfo{howpublished}{\url{https://www.facebook.com/help/447613741984126}}.
\newblock
\newblock
\shownote{Accessed: September 10, 2024}.


\bibitem[Amanda~Lenhart(2024)]%
        {datasociety24}
\bibfield{author}{\bibinfo{person}{Kathryn Zickuhr Myeshia Price-Feeney Amanda~Lenhart, Michele~Ybarra}.} \bibinfo{year}{2024}\natexlab{}.
\newblock \bibinfo{title}{Online Harassment, Digital Abuse, and Cyberstalking}.
\newblock \bibinfo{howpublished}{\url{https://datasociety.net/library/online-harassment-digital-abuse-cyberstalking/}}.
\newblock
\newblock
\shownote{Accessed: September 10, 2024}.


\bibitem[Bennett and Keyes(2020)]%
        {bennett20what}
\bibfield{author}{\bibinfo{person}{Cynthia~L. Bennett} {and} \bibinfo{person}{Os Keyes}.} \bibinfo{year}{2020}\natexlab{}.
\newblock \showarticletitle{What is the point of fairness? disability, AI and the complexity of justice}.
\newblock \bibinfo{journal}{\emph{SIGACCESS Access. Comput.}} \bibinfo{number}{125}, Article \bibinfo{articleno}{5} (\bibinfo{date}{mar} \bibinfo{year}{2020}), \bibinfo{numpages}{1}~pages.
\newblock
\showISSN{1558-2337}
\urldef\tempurl%
\url{https://doi.org/10.1145/3386296.3386301}
\showDOI{\tempurl}


\bibitem[Bertaglia et~al\mbox{.}(2024)]%
        {bertaglia24the}
\bibfield{author}{\bibinfo{person}{Thales Bertaglia}, \bibinfo{person}{Catalina Goanta}, {and} \bibinfo{person}{Adriana Iamnitchi}.} \bibinfo{year}{2024}\natexlab{}.
\newblock \showarticletitle{The Monetisation of Toxicity: Analysing YouTube Content Creators and Controversy-Driven Engagement}. In \bibinfo{booktitle}{\emph{Proceedings of the 4th International Workshop on Open Challenges in Online Social Networks}} (Poznan, Poland) \emph{(\bibinfo{series}{OASIS '24})}. \bibinfo{publisher}{Association for Computing Machinery}, \bibinfo{address}{New York, NY, USA}, \bibinfo{pages}{1–9}.
\newblock
\showISBNx{9798400710827}
\urldef\tempurl%
\url{https://doi.org/10.1145/3677117.3685005}
\showDOI{\tempurl}


\bibitem[Blaser and Ladner(2020)]%
        {blaser20why}
\bibfield{author}{\bibinfo{person}{Brianna Blaser} {and} \bibinfo{person}{Richard~E. Ladner}.} \bibinfo{year}{2020}\natexlab{}.
\newblock \showarticletitle{Why is Data on Disability so Hard to Collect and Understand?}. In \bibinfo{booktitle}{\emph{2020 Research on Equity and Sustained Participation in Engineering, Computing, and Technology (RESPECT)}}, Vol.~\bibinfo{volume}{1}. \bibinfo{pages}{1--8}.
\newblock
\urldef\tempurl%
\url{https://doi.org/10.1109/RESPECT49803.2020.9272466}
\showDOI{\tempurl}


\bibitem[Borgos-Rodriguez(2021)]%
        {borgos-rodriguez21understanding}
\bibfield{author}{\bibinfo{person}{Katya Borgos-Rodriguez}.} \bibinfo{year}{2021}\natexlab{}.
\newblock \showarticletitle{Understanding and Amplifying Labor among Content Creators with Disabilities}. In \bibinfo{booktitle}{\emph{Companion Publication of the 2021 Conference on Computer Supported Cooperative Work and Social Computing}} (Virtual Event, USA) \emph{(\bibinfo{series}{CSCW '21 Companion})}. \bibinfo{publisher}{Association for Computing Machinery}, \bibinfo{address}{New York, NY, USA}, \bibinfo{pages}{241–244}.
\newblock
\showISBNx{9781450384797}
\urldef\tempurl%
\url{https://doi.org/10.1145/3462204.3481784}
\showDOI{\tempurl}


\bibitem[Braun and Clarke(2006)]%
        {braun06using}
\bibfield{author}{\bibinfo{person}{V. Braun} {and} \bibinfo{person}{V. Clarke}.} \bibinfo{year}{2006}\natexlab{}.
\newblock \showarticletitle{Using Thematic Analysis in Psychology}.
\newblock \bibinfo{journal}{\emph{Qualitative Research in Psychology}} \bibinfo{volume}{3}, \bibinfo{number}{2} (\bibinfo{year}{2006}), \bibinfo{pages}{77--101}.
\newblock
\urldef\tempurl%
\url{https://doi.org/10.1191/1478088706qp063oa}
\showDOI{\tempurl}


\bibitem[Bridgland et~al\mbox{.}(2024)]%
        {bridgland24meta}
\bibfield{author}{\bibinfo{person}{Victoria M.~E. Bridgland}, \bibinfo{person}{Payton~J. Jones}, {and} \bibinfo{person}{Benjamin~W. Bellet}.} \bibinfo{year}{2024}\natexlab{}.
\newblock \showarticletitle{A Meta-Analysis of the Efficacy of Trigger Warnings, Content Warnings, and Content Notes}.
\newblock \bibinfo{journal}{\emph{Clinical Psychological Science}} \bibinfo{volume}{12}, \bibinfo{number}{4} (\bibinfo{year}{2024}), \bibinfo{pages}{751--771}.
\newblock
\urldef\tempurl%
\url{https://doi.org/10.1177/21677026231186625}
\showDOI{\tempurl}
\showeprint{https://doi.org/10.1177/21677026231186625}


\bibitem[Campbell(2008)]%
        {campbell08refusing}
\bibfield{author}{\bibinfo{person}{F.~K. Campbell}.} \bibinfo{year}{2008}\natexlab{}.
\newblock \showarticletitle{Refusing Able(ness): A Preliminary Conversation about Ableism}.
\newblock \bibinfo{journal}{\emph{M/C Journal}} \bibinfo{volume}{11}, \bibinfo{number}{3} (\bibinfo{year}{2008}).
\newblock
\urldef\tempurl%
\url{https://doi.org/10.5204/mcj.46}
\showDOI{\tempurl}


\bibitem[Campbell(2010)]%
        {campbell10contours}
\bibfield{author}{\bibinfo{person}{Fiona~Kumari Campbell}.} \bibinfo{year}{2010}\natexlab{}.
\newblock \showarticletitle{Contours of Ableism: The Production of Disability and Abledness}.
\newblock  (\bibinfo{date}{01} \bibinfo{year}{2010}).
\newblock
\showISBNx{978-1-349-36790-0}
\urldef\tempurl%
\url{https://doi.org/10.1057/9780230245181}
\showDOI{\tempurl}


\bibitem[Choi et~al\mbox{.}(2022)]%
        {choi22its}
\bibfield{author}{\bibinfo{person}{Dasom Choi}, \bibinfo{person}{Uichin Lee}, {and} \bibinfo{person}{Hwajung Hong}.} \bibinfo{year}{2022}\natexlab{}.
\newblock \showarticletitle{“It’s not wrong, but I’m quite disappointed”: Toward an Inclusive Algorithmic Experience for Content Creators with Disabilities}. In \bibinfo{booktitle}{\emph{Proceedings of the 2022 CHI Conference on Human Factors in Computing Systems}} (New Orleans, LA, USA) \emph{(\bibinfo{series}{CHI '22})}. \bibinfo{publisher}{Association for Computing Machinery}, \bibinfo{address}{New York, NY, USA}, Article \bibinfo{articleno}{593}, \bibinfo{numpages}{19}~pages.
\newblock
\showISBNx{9781450391573}
\urldef\tempurl%
\url{https://doi.org/10.1145/3491102.3517574}
\showDOI{\tempurl}


\bibitem[Cresci et~al\mbox{.}(2022)]%
        {cresci22personalized}
\bibfield{author}{\bibinfo{person}{Stefano Cresci}, \bibinfo{person}{Amaury Trujillo}, {and} \bibinfo{person}{Tiziano Fagni}.} \bibinfo{year}{2022}\natexlab{}.
\newblock \showarticletitle{Personalized Interventions for Online Moderation}. In \bibinfo{booktitle}{\emph{Proceedings of the 33rd ACM Conference on Hypertext and Social Media}} (Barcelona, Spain) \emph{(\bibinfo{series}{HT '22})}. \bibinfo{publisher}{Association for Computing Machinery}, \bibinfo{address}{New York, NY, USA}, \bibinfo{pages}{248–251}.
\newblock
\showISBNx{9781450392334}
\urldef\tempurl%
\url{https://doi.org/10.1145/3511095.3536369}
\showDOI{\tempurl}


\bibitem[Duggan(2017)]%
        {duggan17online}
\bibfield{author}{\bibinfo{person}{Maeve Duggan}.} \bibinfo{year}{2017}\natexlab{}.
\newblock \showarticletitle{Online Harassment 2017}.
\newblock \bibinfo{journal}{\emph{Pew Research Center}} (\bibinfo{year}{2017}).
\newblock
\urldef\tempurl%
\url{https://www.pewresearch.org/internet/2017/07/11/online-harassment-2017/}
\showURL{%
\tempurl}


\bibitem[Dunn(2021)]%
        {dunn21understanding}
\bibfield{author}{\bibinfo{person}{Dana Dunn}.} \bibinfo{year}{2021}\natexlab{}.
\newblock \bibinfo{title}{Understanding ableism and negative reactions to disability}.
\newblock
\newblock
\urldef\tempurl%
\url{https://www.apa.org/ed/precollege/psychology-teacher-network/introductory-psychology/ableism-negative-reactions-disability}
\showURL{%
\tempurl}


\bibitem[Dunn(2023)]%
        {dunn23avoid}
\bibfield{author}{\bibinfo{person}{Samantha Dunn}.} \bibinfo{year}{2023}\natexlab{}.
\newblock \showarticletitle{Blacklist \& Whitelist: Terms To Avoid}.
\newblock \bibinfo{journal}{\emph{Splunk}} (\bibinfo{year}{2023}).
\newblock
\urldef\tempurl%
\url{https://www.splunk.com/en_us/blog/learn/blacklist-whitelist-inclusivity.html}
\showURL{%
\tempurl}


\bibitem[Duval et~al\mbox{.}(2021)]%
        {duval21chasing}
\bibfield{author}{\bibinfo{person}{Jared Duval}, \bibinfo{person}{Ferran Altarriba~Bertran}, \bibinfo{person}{Siying Chen}, \bibinfo{person}{Melissa Chu}, \bibinfo{person}{Divya Subramonian}, \bibinfo{person}{Austin Wang}, \bibinfo{person}{Geoffrey Xiang}, \bibinfo{person}{Sri Kurniawan}, {and} \bibinfo{person}{Katherine Isbister}.} \bibinfo{year}{2021}\natexlab{}.
\newblock \showarticletitle{Chasing Play on TikTok from Populations with Disabilities to Inspire Playful and Inclusive Technology Design}. In \bibinfo{booktitle}{\emph{Proceedings of the 2021 CHI Conference on Human Factors in Computing Systems}} (Yokohama, Japan) \emph{(\bibinfo{series}{CHI '21})}. \bibinfo{publisher}{Association for Computing Machinery}, \bibinfo{address}{New York, NY, USA}, Article \bibinfo{articleno}{492}, \bibinfo{numpages}{15}~pages.
\newblock
\showISBNx{9781450380966}
\urldef\tempurl%
\url{https://doi.org/10.1145/3411764.3445303}
\showDOI{\tempurl}


\bibitem[Eagle and Ringland(2023)]%
        {eagle23you}
\bibfield{author}{\bibinfo{person}{Tessa Eagle} {and} \bibinfo{person}{Kathryn~E. Ringland}.} \bibinfo{year}{2023}\natexlab{}.
\newblock \showarticletitle{“You Can't Possibly Have ADHD”: Exploring Validation and Tensions around Diagnosis within Unbounded ADHD Social Media Communities}. In \bibinfo{booktitle}{\emph{Proceedings of the 25th International ACM SIGACCESS Conference on Computers and Accessibility}} \emph{(\bibinfo{series}{ASSETS '23})}. \bibinfo{publisher}{Association for Computing Machinery}, \bibinfo{address}{New York, NY, USA}, Article \bibinfo{articleno}{29}, \bibinfo{numpages}{17}~pages.
\newblock
\showISBNx{9798400702204}
\urldef\tempurl%
\url{https://doi.org/10.1145/3597638.3608400}
\showDOI{\tempurl}


\bibitem[Engler(2022)]%
        {engler22middleware}
\bibfield{author}{\bibinfo{person}{Maggie Engler}.} \bibinfo{year}{2022}\natexlab{}.
\newblock \showarticletitle{Middleware and the Customization of Content Moderation}.
\newblock  (\bibinfo{year}{2022}).
\newblock
\urldef\tempurl%
\url{https://integrityinstitute.org/blog/middleware-and-the-customization}
\showURL{%
\tempurl}


\bibitem[Fiesler et~al\mbox{.}(2018)]%
        {fiesler18reddit}
\bibfield{author}{\bibinfo{person}{Casey Fiesler}, \bibinfo{person}{Jialun Jiang}, \bibinfo{person}{Joshua McCann}, \bibinfo{person}{Kyle Frye}, {and} \bibinfo{person}{Jed Brubaker}.} \bibinfo{year}{2018}\natexlab{}.
\newblock \showarticletitle{Reddit Rules! Characterizing an Ecosystem of Governance}.
\newblock \bibinfo{journal}{\emph{Proceedings of the International AAAI Conference on Web and Social Media}} \bibinfo{volume}{12}, \bibinfo{number}{1} (\bibinfo{date}{Jun.} \bibinfo{year}{2018}).
\newblock
\urldef\tempurl%
\url{https://doi.org/10.1609/icwsm.v12i1.15033}
\showDOI{\tempurl}


\bibitem[Francis~Fukuyama({[n.\,d.]})]%
        {fukuyama}
\bibfield{author}{\bibinfo{person}{Ashish Goel Roberta R. Katz A. Douglas Melamed Marietje~Schaake Francis~Fukuyama, Barak~Richman}.} \bibinfo{year}{[n.\,d.]}\natexlab{}.
\newblock \showarticletitle{MIDDLEWARE FOR DOMINANT DIGITAL PLATFORMS: A TECHNOLOGICAL SOLUTION TO A THREAT TO DEMOCRACY}.
\newblock \bibinfo{journal}{\emph{Stanford Cyber Policy Center}} (\bibinfo{year}{[n.\,d.]}).
\newblock


\bibitem[Friedman and Owen(2017)]%
        {friedman17defining}
\bibfield{author}{\bibinfo{person}{Carli Friedman} {and} \bibinfo{person}{Aleksa Owen}.} \bibinfo{year}{2017}\natexlab{}.
\newblock \showarticletitle{Defining Disability: Understandings of and Attitudes Towards Ableism and Disability}.
\newblock \bibinfo{journal}{\emph{Disability Studies Quarterly}}  \bibinfo{volume}{37} (\bibinfo{year}{2017}).
\newblock
\urldef\tempurl%
\url{https://api.semanticscholar.org/CorpusID:151902189}
\showURL{%
\tempurl}


\bibitem[Garg et~al\mbox{.}(2023)]%
        {garg23handling}
\bibfield{author}{\bibinfo{person}{Tanmay Garg}, \bibinfo{person}{Sarah Masud}, \bibinfo{person}{Tharun Suresh}, {and} \bibinfo{person}{Tanmoy Chakraborty}.} \bibinfo{year}{2023}\natexlab{}.
\newblock \showarticletitle{Handling Bias in Toxic Speech Detection: A Survey}.
\newblock \bibinfo{journal}{\emph{ACM Comput. Surv.}} \bibinfo{volume}{55}, \bibinfo{number}{13s}, Article \bibinfo{articleno}{264} (\bibinfo{date}{jul} \bibinfo{year}{2023}), \bibinfo{numpages}{32}~pages.
\newblock
\showISSN{0360-0300}
\urldef\tempurl%
\url{https://doi.org/10.1145/3580494}
\showDOI{\tempurl}


\bibitem[Gillespie(2018)]%
        {gillespie18custodian}
\bibfield{author}{\bibinfo{person}{Tarleton Gillespie}.} \bibinfo{year}{2018}\natexlab{}.
\newblock \bibinfo{booktitle}{\emph{Custodians of the Internet: Platforms, Content Moderation, and the Hidden Decisions That Shape Social Media}}.
\newblock 1--288 pages.
\newblock
\showISBNx{9780300235029}
\urldef\tempurl%
\url{https://doi.org/10.12987/9780300235029}
\showDOI{\tempurl}


\bibitem[Gillespie(2020)]%
        {gillespie20content}
\bibfield{author}{\bibinfo{person}{Tarleton Gillespie}.} \bibinfo{year}{2020}\natexlab{}.
\newblock \showarticletitle{Content moderation, AI, and the question of scale}.
\newblock \bibinfo{journal}{\emph{Big Data \& Society}} \bibinfo{volume}{7}, \bibinfo{number}{2} (\bibinfo{year}{2020}), \bibinfo{pages}{2053951720943234}.
\newblock
\urldef\tempurl%
\url{https://doi.org/10.1177/2053951720943234}
\showDOI{\tempurl}
\showeprint{https://doi.org/10.1177/2053951720943234}


\bibitem[Goyal et~al\mbox{.}(2022)]%
        {goyal22is}
\bibfield{author}{\bibinfo{person}{Nitesh Goyal}, \bibinfo{person}{Ian~D. Kivlichan}, \bibinfo{person}{Rachel Rosen}, {and} \bibinfo{person}{Lucy Vasserman}.} \bibinfo{year}{2022}\natexlab{}.
\newblock \showarticletitle{Is Your Toxicity My Toxicity? Exploring the Impact of Rater Identity on Toxicity Annotation}.
\newblock \bibinfo{journal}{\emph{Proc. ACM Hum.-Comput. Interact.}} \bibinfo{volume}{6}, \bibinfo{number}{CSCW2}, Article \bibinfo{articleno}{363} (\bibinfo{date}{nov} \bibinfo{year}{2022}), \bibinfo{numpages}{28}~pages.
\newblock
\urldef\tempurl%
\url{https://doi.org/10.1145/3555088}
\showDOI{\tempurl}


\bibitem[Grimmelmann(2015)]%
        {grimmelmann15virtues}
\bibfield{author}{\bibinfo{person}{James Grimmelmann}.} \bibinfo{year}{2015}\natexlab{}.
\newblock \showarticletitle{The Virtues of Moderation}.
\newblock \bibinfo{journal}{\emph{Yale Journal of Law \& Technology}}  \bibinfo{volume}{17} (\bibinfo{year}{2015}), \bibinfo{pages}{42}.
\newblock


\bibitem[Haimson et~al\mbox{.}(2020)]%
        {haimson20trans}
\bibfield{author}{\bibinfo{person}{Oliver~L. Haimson}, \bibinfo{person}{Justin Buss}, \bibinfo{person}{Zu Weinger}, \bibinfo{person}{Denny~L. Starks}, \bibinfo{person}{Dykee Gorrell}, {and} \bibinfo{person}{Briar~Sweetbriar Baron}.} \bibinfo{year}{2020}\natexlab{}.
\newblock \showarticletitle{Trans Time: Safety, Privacy, and Content Warnings on a Transgender-Specific Social Media Site}.
\newblock \bibinfo{journal}{\emph{Proc. ACM Hum.-Comput. Interact.}} \bibinfo{volume}{4}, \bibinfo{number}{CSCW2}, Article \bibinfo{articleno}{124} (\bibinfo{date}{oct} \bibinfo{year}{2020}), \bibinfo{numpages}{27}~pages.
\newblock
\urldef\tempurl%
\url{https://doi.org/10.1145/3415195}
\showDOI{\tempurl}


\bibitem[Hakim and Spitzer(2000)]%
        {hakim00effective}
\bibfield{author}{\bibinfo{person}{Jack Hakim} {and} \bibinfo{person}{Tom Spitzer}.} \bibinfo{year}{2000}\natexlab{}.
\newblock \showarticletitle{Effective prototyping for usability}. In \bibinfo{booktitle}{\emph{Proceedings of IEEE Professional Communication Society International Professional Communication Conference and Proceedings of the 18th Annual ACM International Conference on Computer Documentation: Technology \& Teamwork}} (Cambridge, Massachusetts) \emph{(\bibinfo{series}{IPCC/SIGDOC '00})}. \bibinfo{publisher}{IEEE Educational Activities Department}, \bibinfo{address}{USA}, \bibinfo{pages}{47–54}.
\newblock
\showISBNx{0780364317}


\bibitem[Han et~al\mbox{.}(2023)]%
        {han23hate}
\bibfield{author}{\bibinfo{person}{Catherine Han}, \bibinfo{person}{Joseph Seering}, \bibinfo{person}{Deepak Kumar}, \bibinfo{person}{Jeffrey~T. Hancock}, {and} \bibinfo{person}{Zakir Durumeric}.} \bibinfo{year}{2023}\natexlab{}.
\newblock \showarticletitle{Hate Raids on Twitch: Echoes of the Past, New Modalities, and Implications for Platform Governance}.
\newblock \bibinfo{journal}{\emph{Proc. ACM Hum.-Comput. Interact.}} \bibinfo{volume}{7}, \bibinfo{number}{CSCW1}, Article \bibinfo{articleno}{133} (\bibinfo{date}{apr} \bibinfo{year}{2023}), \bibinfo{numpages}{28}~pages.
\newblock
\urldef\tempurl%
\url{https://doi.org/10.1145/3579609}
\showDOI{\tempurl}


\bibitem[Harris et~al\mbox{.}(2023)]%
        {harris23honestly}
\bibfield{author}{\bibinfo{person}{Camille Harris}, \bibinfo{person}{Amber~Gayle Johnson}, \bibinfo{person}{Sadie Palmer}, \bibinfo{person}{Diyi Yang}, {and} \bibinfo{person}{Amy Bruckman}.} \bibinfo{year}{2023}\natexlab{}.
\newblock \showarticletitle{"Honestly, I Think TikTok has a Vendetta Against Black Creators": Understanding Black Content Creator Experiences on TikTok}.
\newblock \bibinfo{journal}{\emph{Proc. ACM Hum.-Comput. Interact.}} \bibinfo{volume}{7}, \bibinfo{number}{CSCW2}, Article \bibinfo{articleno}{320} (\bibinfo{date}{oct} \bibinfo{year}{2023}), \bibinfo{numpages}{31}~pages.
\newblock
\urldef\tempurl%
\url{https://doi.org/10.1145/3610169}
\showDOI{\tempurl}


\bibitem[Hassan et~al\mbox{.}(2021)]%
        {Hassan21}
\bibfield{author}{\bibinfo{person}{Saad Hassan}, \bibinfo{person}{Matt Huenerfauth}, {and} \bibinfo{person}{Cecilia~Ovesdotter Alm}.} \bibinfo{year}{2021}\natexlab{}.
\newblock \showarticletitle{Unpacking the Interdependent Systems of Discrimination: Ableist Bias in NLP Systems through an Intersectional Lens}.
\newblock \bibinfo{journal}{\emph{ArXiv}}  \bibinfo{volume}{abs/2110.00521} (\bibinfo{year}{2021}).
\newblock
\urldef\tempurl%
\url{https://api.semanticscholar.org/CorpusID:238253456}
\showURL{%
\tempurl}


\bibitem[Hehir(2007)]%
        {hehir07confronting}
\bibfield{author}{\bibinfo{person}{Thomas Hehir}.} \bibinfo{year}{2007}\natexlab{}.
\newblock \showarticletitle{Confronting Ableism}.
\newblock \bibinfo{journal}{\emph{Educational Leadership}} (\bibinfo{date}{01} \bibinfo{year}{2007}).
\newblock


\bibitem[Heung et~al\mbox{.}(2024)]%
        {heung24vulnerable}
\bibfield{author}{\bibinfo{person}{Sharon Heung}, \bibinfo{person}{Lucy Jiang}, \bibinfo{person}{Shiri Azenkot}, {and} \bibinfo{person}{Aditya Vashistha}.} \bibinfo{year}{2024}\natexlab{}.
\newblock \showarticletitle{“Vulnerable, Victimized, and Objectified”: Understanding Ableist Hate and Harassment Experienced by Disabled Content Creators on Social Media}. In \bibinfo{booktitle}{\emph{Proceedings of the CHI Conference on Human Factors in Computing Systems}} (Honolulu, HI, USA) \emph{(\bibinfo{series}{CHI '24})}. \bibinfo{publisher}{Association for Computing Machinery}, \bibinfo{address}{New York, NY, USA}, Article \bibinfo{articleno}{744}, \bibinfo{numpages}{19}~pages.
\newblock
\showISBNx{9798400703300}
\urldef\tempurl%
\url{https://doi.org/10.1145/3613904.3641949}
\showDOI{\tempurl}


\bibitem[Heung et~al\mbox{.}(2022)]%
        {heung22nothing}
\bibfield{author}{\bibinfo{person}{Sharon Heung}, \bibinfo{person}{Mahika Phutane}, \bibinfo{person}{Shiri Azenkot}, \bibinfo{person}{Megh Marathe}, {and} \bibinfo{person}{Aditya Vashistha}.} \bibinfo{year}{2022}\natexlab{}.
\newblock \showarticletitle{Nothing Micro About It: Examining Ableist Microaggressions on Social Media}. In \bibinfo{booktitle}{\emph{Proceedings of the 24th International ACM SIGACCESS Conference on Computers and Accessibility}} (Athens, Greece) \emph{(\bibinfo{series}{ASSETS '22})}. \bibinfo{publisher}{Association for Computing Machinery}, \bibinfo{address}{New York, NY, USA}, Article \bibinfo{articleno}{27}, \bibinfo{numpages}{14}~pages.
\newblock
\showISBNx{9781450392587}
\urldef\tempurl%
\url{https://doi.org/10.1145/3517428.3544801}
\showDOI{\tempurl}


\bibitem[Hirsch(2020)]%
        {hirsch20practicing}
\bibfield{author}{\bibinfo{person}{Tad Hirsch}.} \bibinfo{year}{2020}\natexlab{}.
\newblock \showarticletitle{Practicing Without a License: Design Research as Psychotherapy}. In \bibinfo{booktitle}{\emph{Proceedings of the 2020 CHI Conference on Human Factors in Computing Systems}} (Honolulu, HI, USA) \emph{(\bibinfo{series}{CHI '20})}. \bibinfo{publisher}{Association for Computing Machinery}, \bibinfo{address}{New York, NY, USA}, \bibinfo{pages}{1–11}.
\newblock
\showISBNx{9781450367080}
\urldef\tempurl%
\url{https://doi.org/10.1145/3313831.3376750}
\showDOI{\tempurl}


\bibitem[Jhaver et~al\mbox{.}(2022)]%
        {jhaver22designing}
\bibfield{author}{\bibinfo{person}{Shagun Jhaver}, \bibinfo{person}{Quan~Ze Chen}, \bibinfo{person}{Detlef Knauss}, {and} \bibinfo{person}{Amy~X. Zhang}.} \bibinfo{year}{2022}\natexlab{}.
\newblock \showarticletitle{Designing Word Filter Tools for Creator-led Comment Moderation}. In \bibinfo{booktitle}{\emph{Proceedings of the 2022 CHI Conference on Human Factors in Computing Systems}} (New Orleans, LA, USA) \emph{(\bibinfo{series}{CHI '22})}. \bibinfo{publisher}{Association for Computing Machinery}, \bibinfo{address}{New York, NY, USA}, Article \bibinfo{articleno}{205}, \bibinfo{numpages}{21}~pages.
\newblock
\showISBNx{9781450391573}
\urldef\tempurl%
\url{https://doi.org/10.1145/3491102.3517505}
\showDOI{\tempurl}


\bibitem[Jhaver et~al\mbox{.}(2018)]%
        {jhaver18online}
\bibfield{author}{\bibinfo{person}{Shagun Jhaver}, \bibinfo{person}{Sucheta Ghoshal}, \bibinfo{person}{Amy Bruckman}, {and} \bibinfo{person}{Eric Gilbert}.} \bibinfo{year}{2018}\natexlab{}.
\newblock \showarticletitle{Online Harassment and Content Moderation: The Case of Blocklists}.
\newblock \bibinfo{journal}{\emph{ACM Trans. Comput.-Hum. Interact.}} \bibinfo{volume}{25}, \bibinfo{number}{2}, Article \bibinfo{articleno}{12} (\bibinfo{date}{mar} \bibinfo{year}{2018}), \bibinfo{numpages}{33}~pages.
\newblock
\showISSN{1073-0516}
\urldef\tempurl%
\url{https://doi.org/10.1145/3185593}
\showDOI{\tempurl}


\bibitem[Jhaver et~al\mbox{.}(2023)]%
        {jhaver23personalizing}
\bibfield{author}{\bibinfo{person}{Shagun Jhaver}, \bibinfo{person}{Alice~Qian Zhang}, \bibinfo{person}{Quan~Ze Chen}, \bibinfo{person}{Nikhila Natarajan}, \bibinfo{person}{Ruotong Wang}, {and} \bibinfo{person}{Amy~X. Zhang}.} \bibinfo{year}{2023}\natexlab{}.
\newblock \showarticletitle{Personalizing Content Moderation on Social Media: User Perspectives on Moderation Choices, Interface Design, and Labor}.
\newblock \bibinfo{journal}{\emph{Proc. ACM Hum.-Comput. Interact.}} \bibinfo{volume}{7}, \bibinfo{number}{CSCW2}, Article \bibinfo{articleno}{289} (\bibinfo{date}{oct} \bibinfo{year}{2023}), \bibinfo{numpages}{33}~pages.
\newblock
\urldef\tempurl%
\url{https://doi.org/10.1145/3610080}
\showDOI{\tempurl}


\bibitem[Jhaver and Zhang(2023)]%
        {jhaver23do}
\bibfield{author}{\bibinfo{person}{Shagun Jhaver} {and} \bibinfo{person}{Amy~X. Zhang}.} \bibinfo{year}{2023}\natexlab{}.
\newblock \showarticletitle{Do users want platform moderation or individual control? Examining the role of third-person effects and free speech support in shaping moderation preferences}.
\newblock \bibinfo{journal}{\emph{New Media \& Society}} \bibinfo{volume}{0}, \bibinfo{number}{0} (\bibinfo{year}{2023}), \bibinfo{pages}{14614448231217993}.
\newblock
\urldef\tempurl%
\url{https://doi.org/10.1177/14614448231217993}
\showDOI{\tempurl}
\showeprint{https://doi.org/10.1177/14614448231217993}


\bibitem[Jiang et~al\mbox{.}(2023)]%
        {jiang23a}
\bibfield{author}{\bibinfo{person}{Jialun~Aaron Jiang}, \bibinfo{person}{Peipei Nie}, \bibinfo{person}{Jed~R. Brubaker}, {and} \bibinfo{person}{Casey Fiesler}.} \bibinfo{year}{2023}\natexlab{}.
\newblock \showarticletitle{A Trade-off-centered Framework of Content Moderation}.
\newblock \bibinfo{journal}{\emph{ACM Trans. Comput.-Hum. Interact.}} \bibinfo{volume}{30}, \bibinfo{number}{1}, Article \bibinfo{articleno}{3} (\bibinfo{date}{mar} \bibinfo{year}{2023}), \bibinfo{numpages}{34}~pages.
\newblock
\showISSN{1073-0516}
\urldef\tempurl%
\url{https://doi.org/10.1145/3534929}
\showDOI{\tempurl}


\bibitem[Jiang et~al\mbox{.}(2021)]%
        {jiang21understanding}
\bibfield{author}{\bibinfo{person}{Jialun~Aaron Jiang}, \bibinfo{person}{Morgan~Klaus Scheuerman}, \bibinfo{person}{Casey Fiesler}, {and} \bibinfo{person}{Jed~R Brubaker}.} \bibinfo{year}{2021}\natexlab{}.
\newblock \showarticletitle{{Understanding international perceptions of the severity of harmful content online}}.
\newblock \bibinfo{journal}{\emph{PLOS ONE}} \bibinfo{volume}{16}, \bibinfo{number}{8} (\bibinfo{date}{August} \bibinfo{year}{2021}), \bibinfo{pages}{1--22}.
\newblock
\urldef\tempurl%
\url{https://doi.org/10.1371/journal.pone.0256}
\showDOI{\tempurl}


\bibitem[Jigsaw(2019)]%
        {jigsaw19tune}
\bibfield{author}{\bibinfo{person}{Jigsaw}.} \bibinfo{year}{2019}\natexlab{}.
\newblock \bibinfo{title}{{T}une: {C}ontrol the comments you see}.
\newblock
\newblock
\urldef\tempurl%
\url{https://medium.com/jigsaw/tune-control-the-comments-you-see-b10cc807a171}
\showURL{%
\tempurl}
\newblock
\shownote{[Accessed 11-09-2024]}.


\bibitem[Johnson(2019)]%
        {johnson19inclusion}
\bibfield{author}{\bibinfo{person}{Mark~R. Johnson}.} \bibinfo{year}{2019}\natexlab{}.
\newblock \showarticletitle{Inclusion and exclusion in the digital economy: disability and mental health as a live streamer on {Twitch}.tv}.
\newblock \bibinfo{journal}{\emph{Information, Communication \& Society}} \bibinfo{volume}{22}, \bibinfo{number}{4} (\bibinfo{date}{March} \bibinfo{year}{2019}), \bibinfo{pages}{506--520}.
\newblock
\showISSN{1369-118X}
\urldef\tempurl%
\url{https://doi.org/10.1080/1369118X.2018.1476575}
\showDOI{\tempurl}
\newblock
\shownote{Publisher: Routledge \_eprint: https://doi.org/10.1080/1369118X.2018.1476575}.


\bibitem[Karizat et~al\mbox{.}(2021)]%
        {karizat21algorithmic}
\bibfield{author}{\bibinfo{person}{Nadia Karizat}, \bibinfo{person}{Dan Delmonaco}, \bibinfo{person}{Motahhare Eslami}, {and} \bibinfo{person}{Nazanin Andalibi}.} \bibinfo{year}{2021}\natexlab{}.
\newblock \showarticletitle{Algorithmic Folk Theories and Identity: How TikTok Users Co-Produce Knowledge of Identity and Engage in Algorithmic Resistance}.
\newblock \bibinfo{journal}{\emph{Proc. ACM Hum.-Comput. Interact.}} \bibinfo{volume}{5}, \bibinfo{number}{CSCW2}, Article \bibinfo{articleno}{305} (\bibinfo{date}{oct} \bibinfo{year}{2021}), \bibinfo{numpages}{44}~pages.
\newblock
\urldef\tempurl%
\url{https://doi.org/10.1145/3476046}
\showDOI{\tempurl}


\bibitem[Kaur et~al\mbox{.}(2024)]%
        {kaur24challenges}
\bibfield{author}{\bibinfo{person}{Sukhnidh Kaur}, \bibinfo{person}{Manohar Swaminathan}, \bibinfo{person}{Kalika Bali}, {and} \bibinfo{person}{Aditya Vashistha}.} \bibinfo{year}{2024}\natexlab{}.
\newblock \showarticletitle{Challenges to Online Disability Rights Advocacy in India}. In \bibinfo{booktitle}{\emph{Proceedings of the CHI Conference on Human Factors in Computing Systems}} (Honolulu, HI, USA) \emph{(\bibinfo{series}{CHI '24})}. \bibinfo{publisher}{Association for Computing Machinery}, \bibinfo{address}{New York, NY, USA}, Article \bibinfo{articleno}{397}, \bibinfo{numpages}{15}~pages.
\newblock
\showISBNx{9798400703300}
\urldef\tempurl%
\url{https://doi.org/10.1145/3613904.3642737}
\showDOI{\tempurl}


\bibitem[Keller and Galgay(2010)]%
        {keller10microagggressions}
\bibfield{author}{\bibinfo{person}{R.M. Keller} {and} \bibinfo{person}{Corinne Galgay}.} \bibinfo{year}{2010}\natexlab{}.
\newblock \showarticletitle{Microaggressions experienced by people with disabilities in US society}.
\newblock \bibinfo{journal}{\emph{Microaggressions and marginality: Manifestation, dynamics, and impact}} (\bibinfo{date}{01} \bibinfo{year}{2010}), \bibinfo{pages}{241--268}.
\newblock


\bibitem[Lee and Rich(2021)]%
        {lee21who}
\bibfield{author}{\bibinfo{person}{Min~Kyung Lee} {and} \bibinfo{person}{Katherine Rich}.} \bibinfo{year}{2021}\natexlab{}.
\newblock \showarticletitle{Who Is Included in Human Perceptions of AI?: Trust and Perceived Fairness around Healthcare AI and Cultural Mistrust}. In \bibinfo{booktitle}{\emph{Proceedings of the 2021 CHI Conference on Human Factors in Computing Systems}} (Yokohama, Japan) \emph{(\bibinfo{series}{CHI '21})}. \bibinfo{publisher}{Association for Computing Machinery}, \bibinfo{address}{New York, NY, USA}, Article \bibinfo{articleno}{138}, \bibinfo{numpages}{14}~pages.
\newblock
\showISBNx{9781450380966}
\urldef\tempurl%
\url{https://doi.org/10.1145/3411764.3445570}
\showDOI{\tempurl}


\bibitem[Lewis(2022)]%
        {lewis22working}
\bibfield{author}{\bibinfo{person}{T.~A. Lewis}.} \bibinfo{year}{2022}\natexlab{}.
\newblock \bibinfo{booktitle}{\emph{Working Definition of Ableism - January 2022 Update}}.
\newblock
\urldef\tempurl%
\url{https://www.talilalewis.com/blog/workingdefinition-of-ableism-january-2022-update}
\showURL{%
\tempurl}
\newblock
\shownote{Accessed: 2024-12-03}.


\bibitem[Li et~al\mbox{.}(2024)]%
        {li24chatgpt}
\bibfield{author}{\bibinfo{person}{Lingyao Li}, \bibinfo{person}{Lizhou Fan}, \bibinfo{person}{Shubham Atreja}, {and} \bibinfo{person}{Libby Hemphill}.} \bibinfo{year}{2024}\natexlab{}.
\newblock \showarticletitle{“HOT” ChatGPT: The Promise of ChatGPT in Detecting and Discriminating Hateful, Offensive, and Toxic Comments on Social Media}.
\newblock \bibinfo{journal}{\emph{ACM Trans. Web}} \bibinfo{volume}{18}, \bibinfo{number}{2}, Article \bibinfo{articleno}{30} (\bibinfo{date}{mar} \bibinfo{year}{2024}), \bibinfo{numpages}{36}~pages.
\newblock
\showISSN{1559-1131}
\urldef\tempurl%
\url{https://doi.org/10.1145/3643829}
\showDOI{\tempurl}


\bibitem[Lyu and Carroll(2024)]%
        {lyu24because}
\bibfield{author}{\bibinfo{person}{Yao Lyu} {and} \bibinfo{person}{John~M. Carroll}.} \bibinfo{year}{2024}\natexlab{}.
\newblock \showarticletitle{"Because Some Sighted People, They Don't Know What the Heck You're Talking About:" A Study of Blind Tokers' Infrastructuring Work to Build Independence}.
\newblock \bibinfo{journal}{\emph{Proc. ACM Hum.-Comput. Interact.}} \bibinfo{volume}{8}, \bibinfo{number}{CSCW1}, Article \bibinfo{articleno}{20} (\bibinfo{date}{apr} \bibinfo{year}{2024}), \bibinfo{numpages}{30}~pages.
\newblock
\urldef\tempurl%
\url{https://doi.org/10.1145/3637297}
\showDOI{\tempurl}


\bibitem[Ma and Kou(2023)]%
        {ma23defaulting}
\bibfield{author}{\bibinfo{person}{Renkai Ma} {and} \bibinfo{person}{Yubo Kou}.} \bibinfo{year}{2023}\natexlab{}.
\newblock \showarticletitle{"Defaulting to boilerplate answers, they didn't engage in a genuine conversation": Dimensions of Transparency Design in Creator Moderation}.
\newblock \bibinfo{journal}{\emph{Proc. ACM Hum.-Comput. Interact.}} \bibinfo{volume}{7}, \bibinfo{number}{CSCW1}, Article \bibinfo{articleno}{44} (\bibinfo{date}{apr} \bibinfo{year}{2023}), \bibinfo{numpages}{26}~pages.
\newblock
\urldef\tempurl%
\url{https://doi.org/10.1145/3579477}
\showDOI{\tempurl}


\bibitem[Maarouf et~al\mbox{.}(2024)]%
        {maarouf24the}
\bibfield{author}{\bibinfo{person}{Abdurahman Maarouf}, \bibinfo{person}{Nicolas Pr\"{o}llochs}, {and} \bibinfo{person}{Stefan Feuerriegel}.} \bibinfo{year}{2024}\natexlab{}.
\newblock \showarticletitle{The Virality of Hate Speech on Social Media}.
\newblock \bibinfo{journal}{\emph{Proc. ACM Hum.-Comput. Interact.}} \bibinfo{volume}{8}, \bibinfo{number}{CSCW1}, Article \bibinfo{articleno}{186} (\bibinfo{date}{apr} \bibinfo{year}{2024}), \bibinfo{numpages}{22}~pages.
\newblock
\urldef\tempurl%
\url{https://doi.org/10.1145/3641025}
\showDOI{\tempurl}


\bibitem[Manne(2015)]%
        {manne25why}
\bibfield{author}{\bibinfo{person}{Kate Manne}.} \bibinfo{year}{2015}\natexlab{}.
\newblock \showarticletitle{Why I Use Trigger Warnings}.
\newblock \bibinfo{journal}{\emph{The New York Times}} (\bibinfo{year}{2015}).
\newblock
\urldef\tempurl%
\url{https://www.nytimes.com/2015/09/20/opinion/sunday/why-i-use-trigger-warnings.html}
\showURL{%
\tempurl}


\bibitem[McGillicuddy et~al\mbox{.}(2016)]%
        {mcgillicuddy2016controlling}
\bibfield{author}{\bibinfo{person}{A McGillicuddy}, \bibinfo{person}{Jean-Gregoire Bernard}, {and} \bibinfo{person}{Jocelyn Cranefield}.} \bibinfo{year}{2016}\natexlab{}.
\newblock \showarticletitle{{Controlling Bad Behavior in Online Communities: An Examination of Moderation Work}}.
\newblock  (\bibinfo{date}{1} \bibinfo{year}{2016}).
\newblock
\urldef\tempurl%
\url{https://doi.org/10.26686/wgtn.12910085.v1}
\showDOI{\tempurl}


\bibitem[Ming et~al\mbox{.}(2021)]%
        {ming21accept}
\bibfield{author}{\bibinfo{person}{Joy Ming}, \bibinfo{person}{Sharon Heung}, \bibinfo{person}{Shiri Azenkot}, {and} \bibinfo{person}{Aditya Vashistha}.} \bibinfo{year}{2021}\natexlab{}.
\newblock \showarticletitle{Accept or Address? Researchers’ Perspectives on Response Bias in Accessibility Research}. In \bibinfo{booktitle}{\emph{Proceedings of the 23rd International ACM SIGACCESS Conference on Computers and Accessibility}} (Virtual Event, USA) \emph{(\bibinfo{series}{ASSETS '21})}. \bibinfo{publisher}{Association for Computing Machinery}, \bibinfo{address}{New York, NY, USA}, Article \bibinfo{articleno}{20}, \bibinfo{numpages}{13}~pages.
\newblock
\showISBNx{9781450383066}
\urldef\tempurl%
\url{https://doi.org/10.1145/3441852.3471216}
\showDOI{\tempurl}


\bibitem[Mok et~al\mbox{.}(2023)]%
        {mok23experiences}
\bibfield{author}{\bibinfo{person}{Terrance Mok}, \bibinfo{person}{Anthony Tang}, \bibinfo{person}{Adam McCrimmon}, {and} \bibinfo{person}{Lora Oehlberg}.} \bibinfo{year}{2023}\natexlab{}.
\newblock \showarticletitle{Experiences of Autistic Twitch Livestreamers: “I Have Made Easily the Most Meaningful and Impactful Relationships”}. In \bibinfo{booktitle}{\emph{Proceedings of the 25th International ACM SIGACCESS Conference on Computers and Accessibility}} \emph{(\bibinfo{series}{ASSETS '23})}. \bibinfo{publisher}{Association for Computing Machinery}, \bibinfo{address}{New York, NY, USA}, Article \bibinfo{articleno}{41}, \bibinfo{numpages}{15}~pages.
\newblock
\showISBNx{9798400702204}
\urldef\tempurl%
\url{https://doi.org/10.1145/3597638.3608416}
\showDOI{\tempurl}


\bibitem[Moolenijzer and Dew(2023)]%
        {moolenijzer23they}
\bibfield{author}{\bibinfo{person}{Nicolaas~B Moolenijzer} {and} \bibinfo{person}{Kristin Dew}.} \bibinfo{year}{2023}\natexlab{}.
\newblock \showarticletitle{“They know that it works because we are looking for ourselves” – LGBTQ+ TikTok Users' Perceptions and Experiences of Queerbaiting}. In \bibinfo{booktitle}{\emph{Proceedings of the 25th International Conference on Mobile Human-Computer Interaction}} (Athens, Greece) \emph{(\bibinfo{series}{MobileHCI '23 Companion})}. \bibinfo{publisher}{Association for Computing Machinery}, \bibinfo{address}{New York, NY, USA}, Article \bibinfo{articleno}{20}, \bibinfo{numpages}{6}~pages.
\newblock
\showISBNx{9781450399241}
\urldef\tempurl%
\url{https://doi.org/10.1145/3565066.3608705}
\showDOI{\tempurl}


\bibitem[Musgrave et~al\mbox{.}(2022)]%
        {musgrave22experiences}
\bibfield{author}{\bibinfo{person}{Tyler Musgrave}, \bibinfo{person}{Alia Cummings}, {and} \bibinfo{person}{Sarita Schoenebeck}.} \bibinfo{year}{2022}\natexlab{}.
\newblock \showarticletitle{Experiences of Harm, Healing, and Joy among Black Women and Femmes on Social Media}. In \bibinfo{booktitle}{\emph{Proceedings of the 2022 CHI Conference on Human Factors in Computing Systems}} (New Orleans, LA, USA) \emph{(\bibinfo{series}{CHI '22})}. \bibinfo{publisher}{Association for Computing Machinery}, \bibinfo{address}{New York, NY, USA}, Article \bibinfo{articleno}{240}, \bibinfo{numpages}{17}~pages.
\newblock
\showISBNx{9781450391573}
\urldef\tempurl%
\url{https://doi.org/10.1145/3491102.3517608}
\showDOI{\tempurl}


\bibitem[Nieva(2024)]%
        {nieva24heres}
\bibfield{author}{\bibinfo{person}{Richard Nieva}.} \bibinfo{year}{2024}\natexlab{}.
\newblock \bibinfo{title}{Here’s How Facebook Uses Artificial Intelligence to Take Down Abusive Posts}.
\newblock \bibinfo{howpublished}{\url{https://www.cnet.com/tech/tech-industry/heres-how-facebook-uses-artificial-intelligence-to-take-down-abusive-posts-f8/}}.
\newblock
\newblock
\shownote{Accessed: September 10, 2024}.


\bibitem[Niu et~al\mbox{.}(2024)]%
        {Niu24please}
\bibfield{author}{\bibinfo{person}{Shuo Niu}, \bibinfo{person}{Li Liu}, {and} \bibinfo{person}{Yali Bian}.} \bibinfo{year}{2024}\natexlab{}.
\newblock \showarticletitle{Please Understand My Disability: An Analysis of YouTubers' Discourse on Disability Challenges}.
\newblock \bibinfo{journal}{\emph{Proc. ACM Hum.-Comput. Interact.}} \bibinfo{volume}{8}, \bibinfo{number}{CSCW2}, Article \bibinfo{articleno}{407} (\bibinfo{date}{Nov.} \bibinfo{year}{2024}), \bibinfo{numpages}{25}~pages.
\newblock
\urldef\tempurl%
\url{https://doi.org/10.1145/3686946}
\showDOI{\tempurl}


\bibitem[Oliva(2021)]%
        {oliva21fighting}
\bibfield{author}{\bibinfo{person}{Thiago Oliva}.} \bibinfo{year}{2021}\natexlab{}.
\newblock \showarticletitle{Fighting Hate Speech, Silencing Drag Queens? Artificial Intelligence in Content Moderation and Risks to LGBTQ Voices Online}.
\newblock \bibinfo{journal}{\emph{Sexuality \& Culture}} (\bibinfo{date}{04} \bibinfo{year}{2021}).
\newblock
\urldef\tempurl%
\url{https://doi.org/10.1007/s12119-020-09790-w}
\showDOI{\tempurl}


\bibitem[Pater et~al\mbox{.}(2016)]%
        {pater16characterizations}
\bibfield{author}{\bibinfo{person}{Jessica~A. Pater}, \bibinfo{person}{Moon~K. Kim}, \bibinfo{person}{Elizabeth~D. Mynatt}, {and} \bibinfo{person}{Casey Fiesler}.} \bibinfo{year}{2016}\natexlab{}.
\newblock \showarticletitle{Characterizations of Online Harassment: Comparing Policies Across Social Media Platforms}. In \bibinfo{booktitle}{\emph{Proceedings of the 2016 ACM International Conference on Supporting Group Work}} (Sanibel Island, Florida, USA) \emph{(\bibinfo{series}{GROUP '16})}. \bibinfo{publisher}{Association for Computing Machinery}, \bibinfo{address}{New York, NY, USA}, \bibinfo{pages}{369–374}.
\newblock
\showISBNx{9781450342766}
\urldef\tempurl%
\url{https://doi.org/10.1145/2957276.2957297}
\showDOI{\tempurl}


\bibitem[Porter(2021)]%
        {porter21today}
\bibfield{author}{\bibinfo{person}{Jon Porter}.} \bibinfo{year}{2021}\natexlab{}.
\newblock \bibinfo{title}{Today I learned about Intel’s AI sliders that filter online gaming abuse}.
\newblock
\newblock
\urldef\tempurl%
\url{https://www.theverge.com/2021/4/8/22373290/intel-bleep-ai-powered-abuse-toxicity-gaming-filters}
\showURL{%
\tempurl}


\bibitem[Rauchberg(2022)]%
        {rauchberg22shadowbanned}
\bibfield{author}{\bibinfo{person}{Jessiva~Sage Rauchberg}.} \bibinfo{year}{2022}\natexlab{}.
\newblock \bibinfo{booktitle}{\emph{\#Shadowbanned: Queer, Trans, and Disabled Creator Responses to Algorithmic Oppression on TikTok} (\bibinfo{edition}{1st edition} ed.)}.
\newblock 196 -- 210 pages.
\newblock
\showISBNx{9781003196457}


\bibitem[Ringland(2019)]%
        {ringland19autsome}
\bibfield{author}{\bibinfo{person}{Kathryn~E. Ringland}.} \bibinfo{year}{2019}\natexlab{}.
\newblock \showarticletitle{“{Autsome}”: {Fostering} an {Autistic} {Identity} in an {Online} {Minecraft} {Community} for {Youth} with {Autism}}.
\newblock \bibinfo{journal}{\emph{Information in Contemporary Society : 14th International Conference, iConference 2019, Washington, DC, USA, March 31-April 3, 2019, Proceedings. iConference (Conference) (14th : 2019 : Washington, D.C.)}}  \bibinfo{volume}{11420} (\bibinfo{date}{April} \bibinfo{year}{2019}), \bibinfo{pages}{132--143}.
\newblock
\urldef\tempurl%
\url{https://doi.org/10.1007/978-3-030-15742-5_12}
\showDOI{\tempurl}


\bibitem[Ringland et~al\mbox{.}(2019)]%
        {ringland19understanding}
\bibfield{author}{\bibinfo{person}{Kathryn~E. Ringland} {et~al\mbox{.}}} \bibinfo{year}{2019}\natexlab{}.
\newblock \showarticletitle{Understanding Mental Ill-health as Psychosocial Disability: Implications for Assistive Technology}. In \bibinfo{booktitle}{\emph{Proceedings of the Annual ACM Conference on Assistive Technologies (ASSETS)}}, Vol.~\bibinfo{volume}{2019}. \bibinfo{publisher}{ACM}, \bibinfo{pages}{156--170}.
\newblock
\urldef\tempurl%
\url{https://doi.org/10.1145/3308561.3353785}
\showDOI{\tempurl}


\bibitem[Rong et~al\mbox{.}(2022)]%
        {rong22it}
\bibfield{author}{\bibinfo{person}{Ethan~Z. Rong}, \bibinfo{person}{Mo~Morgana Zhou}, \bibinfo{person}{Zhicong Lu}, {and} \bibinfo{person}{Mingming Fan}.} \bibinfo{year}{2022}\natexlab{}.
\newblock \showarticletitle{“It Feels Like Being Locked in A Cage”: Understanding Blind or Low Vision Streamers’ Perceptions of Content Curation Algorithms}. In \bibinfo{booktitle}{\emph{Proceedings of the 2022 ACM Designing Interactive Systems Conference}} (Virtual Event, Australia) \emph{(\bibinfo{series}{DIS '22})}. \bibinfo{publisher}{Association for Computing Machinery}, \bibinfo{address}{New York, NY, USA}, \bibinfo{pages}{571–585}.
\newblock
\showISBNx{9781450393584}
\urldef\tempurl%
\url{https://doi.org/10.1145/3532106.3533514}
\showDOI{\tempurl}


\bibitem[Rudd et~al\mbox{.}(1996)]%
        {rudd96low}
\bibfield{author}{\bibinfo{person}{Jim Rudd}, \bibinfo{person}{Ken Stern}, {and} \bibinfo{person}{Scott Isensee}.} \bibinfo{year}{1996}\natexlab{}.
\newblock \showarticletitle{Low vs. high-fidelity prototyping debate}.
\newblock \bibinfo{journal}{\emph{Interactions}} \bibinfo{volume}{3}, \bibinfo{number}{1} (\bibinfo{date}{jan} \bibinfo{year}{1996}), \bibinfo{pages}{76–85}.
\newblock
\showISSN{1072-5520}
\urldef\tempurl%
\url{https://doi.org/10.1145/223500.223514}
\showDOI{\tempurl}


\bibitem[Samermit et~al\mbox{.}(2023)]%
        {samermit23millions}
\bibfield{author}{\bibinfo{person}{Patrawat Samermit}, \bibinfo{person}{Anna Turner}, \bibinfo{person}{Patrick~Gage Kelley}, \bibinfo{person}{Tara Matthews}, \bibinfo{person}{Vanessia Wu}, \bibinfo{person}{Sunny Consolvo}, {and} \bibinfo{person}{Kurt Thomas}.} \bibinfo{year}{2023}\natexlab{}.
\newblock \showarticletitle{{{\textquotedblleft}Millions} of people are watching {you{\textquotedblright}}: Understanding the {Digital-Safety} Needs and Practices of Creators}. In \bibinfo{booktitle}{\emph{32nd USENIX Security Symposium (USENIX Security 23)}}. \bibinfo{publisher}{USENIX Association}, \bibinfo{address}{Anaheim, CA}, \bibinfo{pages}{5629--5645}.
\newblock
\showISBNx{978-1-939133-37-3}
\urldef\tempurl%
\url{https://www.usenix.org/conference/usenixsecurity23/presentation/samermit}
\showURL{%
\tempurl}


\bibitem[Sannon et~al\mbox{.}(2023)]%
        {sannon23disability}
\bibfield{author}{\bibinfo{person}{Shruti Sannon}, \bibinfo{person}{Jordyn Young}, \bibinfo{person}{Erica Shusas}, {and} \bibinfo{person}{Andrea Forte}.} \bibinfo{year}{2023}\natexlab{}.
\newblock \showarticletitle{Disability Activism on Social Media: Sociotechnical Challenges in the Pursuit of Visibility}. In \bibinfo{booktitle}{\emph{Proceedings of the 2023 CHI Conference on Human Factors in Computing Systems}} (Hamburg, Germany) \emph{(\bibinfo{series}{CHI '23})}. \bibinfo{publisher}{Association for Computing Machinery}, \bibinfo{address}{New York, NY, USA}, Article \bibinfo{articleno}{672}, \bibinfo{numpages}{15}~pages.
\newblock
\showISBNx{9781450394215}
\urldef\tempurl%
\url{https://doi.org/10.1145/3544548.3581333}
\showDOI{\tempurl}


\bibitem[Sap et~al\mbox{.}(2019)]%
        {sap19the}
\bibfield{author}{\bibinfo{person}{Maarten Sap}, \bibinfo{person}{Dallas Card}, \bibinfo{person}{Saadia Gabriel}, \bibinfo{person}{Yejin Choi}, {and} \bibinfo{person}{Noah~A. Smith}.} \bibinfo{year}{2019}\natexlab{}.
\newblock \showarticletitle{The Risk of Racial Bias in Hate Speech Detection}. In \bibinfo{booktitle}{\emph{Proceedings of the 57th Annual Meeting of the Association for Computational Linguistics}}, \bibfield{editor}{\bibinfo{person}{Anna Korhonen}, \bibinfo{person}{David Traum}, {and} \bibinfo{person}{Llu{\'\i}s M{\`a}rquez}} (Eds.). \bibinfo{publisher}{Association for Computational Linguistics}, \bibinfo{address}{Florence, Italy}, \bibinfo{pages}{1668--1678}.
\newblock
\urldef\tempurl%
\url{https://doi.org/10.18653/v1/P19-1163}
\showDOI{\tempurl}


\bibitem[Schafer et~al\mbox{.}(2023)]%
        {schafer23participatory}
\bibfield{author}{\bibinfo{person}{Joseph~S. Schafer}, \bibinfo{person}{Kate Starbird}, {and} \bibinfo{person}{Daniela~K. Rosner}.} \bibinfo{year}{2023}\natexlab{}.
\newblock \showarticletitle{Participatory Design and Power in Misinformation, Disinformation, and Online Hate Research}. In \bibinfo{booktitle}{\emph{Proceedings of the 2023 ACM Designing Interactive Systems Conference}} (Pittsburgh, PA, USA) \emph{(\bibinfo{series}{DIS '23})}. \bibinfo{publisher}{Association for Computing Machinery}, \bibinfo{address}{New York, NY, USA}, \bibinfo{pages}{1724–1739}.
\newblock
\showISBNx{9781450398930}
\urldef\tempurl%
\url{https://doi.org/10.1145/3563657.3596119}
\showDOI{\tempurl}


\bibitem[Scheuerman et~al\mbox{.}(2018)]%
        {scheuerman18safe}
\bibfield{author}{\bibinfo{person}{Morgan~Klaus Scheuerman}, \bibinfo{person}{Stacy~M. Branham}, {and} \bibinfo{person}{Foad Hamidi}.} \bibinfo{year}{2018}\natexlab{}.
\newblock \showarticletitle{Safe Spaces and Safe Places: Unpacking Technology-Mediated Experiences of Safety and Harm with Transgender People}.
\newblock \bibinfo{journal}{\emph{Proc. ACM Hum.-Comput. Interact.}} \bibinfo{volume}{2}, \bibinfo{number}{CSCW}, Article \bibinfo{articleno}{155} (\bibinfo{date}{nov} \bibinfo{year}{2018}), \bibinfo{numpages}{27}~pages.
\newblock
\urldef\tempurl%
\url{https://doi.org/10.1145/3274424}
\showDOI{\tempurl}


\bibitem[Scheuerman et~al\mbox{.}(2021)]%
        {scheuerman21framework}
\bibfield{author}{\bibinfo{person}{Morgan~Klaus Scheuerman}, \bibinfo{person}{Jialun~Aaron Jiang}, \bibinfo{person}{Casey Fiesler}, {and} \bibinfo{person}{Jed~R. Brubaker}.} \bibinfo{year}{2021}\natexlab{}.
\newblock \showarticletitle{A Framework of Severity for Harmful Content Online}.
\newblock \bibinfo{journal}{\emph{Proc. ACM Hum.-Comput. Interact.}} \bibinfo{volume}{5}, \bibinfo{number}{CSCW2}, Article \bibinfo{articleno}{368} (\bibinfo{date}{oct} \bibinfo{year}{2021}), \bibinfo{numpages}{33}~pages.
\newblock
\urldef\tempurl%
\url{https://doi.org/10.1145/3479512}
\showDOI{\tempurl}


\bibitem[Schoenebeck et~al\mbox{.}(2023)]%
        {schoenebeck23online}
\bibfield{author}{\bibinfo{person}{Sarita Schoenebeck}, \bibinfo{person}{Amna Batool}, \bibinfo{person}{Giang Do}, \bibinfo{person}{Sylvia Darling}, \bibinfo{person}{Gabriel Grill}, \bibinfo{person}{Daricia Wilkinson}, \bibinfo{person}{Mehtab Khan}, \bibinfo{person}{Kentaro Toyama}, {and} \bibinfo{person}{Louise Ashwell}.} \bibinfo{year}{2023}\natexlab{}.
\newblock \showarticletitle{Online Harassment in Majority Contexts: Examining Harms and Remedies across Countries}. In \bibinfo{booktitle}{\emph{Proceedings of the 2023 CHI Conference on Human Factors in Computing Systems}} \emph{(\bibinfo{series}{CHI '23})}. \bibinfo{publisher}{Association for Computing Machinery}, \bibinfo{address}{New York, NY, USA}, Article \bibinfo{articleno}{485}, \bibinfo{numpages}{16}~pages.
\newblock
\showISBNx{9781450394215}
\urldef\tempurl%
\url{https://doi.org/10.1145/3544548.3581020}
\showDOI{\tempurl}


\bibitem[Scott et~al\mbox{.}(2023)]%
        {scott23trauma}
\bibfield{author}{\bibinfo{person}{Carol~F Scott}, \bibinfo{person}{Gabriela Marcu}, \bibinfo{person}{Riana~Elyse Anderson}, \bibinfo{person}{Mark~W Newman}, {and} \bibinfo{person}{Sarita Schoenebeck}.} \bibinfo{year}{2023}\natexlab{}.
\newblock \showarticletitle{Trauma-Informed Social Media: Towards Solutions for Reducing and Healing Online Harm}. In \bibinfo{booktitle}{\emph{Proceedings of the 2023 CHI Conference on Human Factors in Computing Systems}} \emph{(\bibinfo{series}{CHI '23})}. \bibinfo{publisher}{Association for Computing Machinery}, \bibinfo{address}{New York, NY, USA}, Article \bibinfo{articleno}{341}, \bibinfo{numpages}{20}~pages.
\newblock
\showISBNx{9781450394215}
\urldef\tempurl%
\url{https://doi.org/10.1145/3544548.3581512}
\showDOI{\tempurl}


\bibitem[Shahid and Vashistha(2023)]%
        {shahid23decolonizing}
\bibfield{author}{\bibinfo{person}{Farhana Shahid} {and} \bibinfo{person}{Aditya Vashistha}.} \bibinfo{year}{2023}\natexlab{}.
\newblock \showarticletitle{Decolonizing Content Moderation: Does Uniform Global Community Standard Resemble Utopian Equality or Western Power Hegemony?}. In \bibinfo{booktitle}{\emph{Proceedings of the 2023 CHI Conference on Human Factors in Computing Systems}} (Hamburg, Germany) \emph{(\bibinfo{series}{CHI '23})}. \bibinfo{publisher}{Association for Computing Machinery}, \bibinfo{address}{New York, NY, USA}, Article \bibinfo{articleno}{391}, \bibinfo{numpages}{18}~pages.
\newblock
\showISBNx{9781450394215}
\urldef\tempurl%
\url{https://doi.org/10.1145/3544548.3581538}
\showDOI{\tempurl}


\bibitem[Simpson et~al\mbox{.}(2023)]%
        {simpson23hey}
\bibfield{author}{\bibinfo{person}{Ellen Simpson}, \bibinfo{person}{Samantha Dalal}, {and} \bibinfo{person}{Bryan Semaan}.} \bibinfo{year}{2023}\natexlab{}.
\newblock \showarticletitle{"Hey, Can You Add Captions?": The Critical Infrastructuring Practices of Neurodiverse People on TikTok}.
\newblock \bibinfo{journal}{\emph{Proc. ACM Hum.-Comput. Interact.}} \bibinfo{volume}{7}, \bibinfo{number}{CSCW1}, Article \bibinfo{articleno}{57} (\bibinfo{date}{apr} \bibinfo{year}{2023}), \bibinfo{numpages}{27}~pages.
\newblock
\urldef\tempurl%
\url{https://doi.org/10.1145/3579490}
\showDOI{\tempurl}


\bibitem[Thomas et~al\mbox{.}(2021)]%
        {thomas21sok}
\bibfield{editor}{\bibinfo{person}{Kurt Thomas}, \bibinfo{person}{Devdatta Akhawe}, \bibinfo{person}{Michael Bailey}, \bibinfo{person}{Dan Boneh}, \bibinfo{person}{Elie Bursztein}, \bibinfo{person}{Sunny Consolvo}, \bibinfo{person}{Nicola Dell}, \bibinfo{person}{Zakir Durumeric}, \bibinfo{person}{Patrick~Gage Kelley}, \bibinfo{person}{Deepak Kumar}, \bibinfo{person}{Damon McCoy}, \bibinfo{person}{Sarah Meiklejohn}, \bibinfo{person}{Thomas Ristenpart}, {and} \bibinfo{person}{Gianluca Stringhini}} (Eds.). \bibinfo{year}{2021}\natexlab{}.
\newblock \bibinfo{booktitle}{\emph{SoK: Hate, Harassment, and the Changing Landscape of Online Abuse}}.
\newblock


\bibitem[Thomas et~al\mbox{.}(2022)]%
        {thomas22its}
\bibfield{editor}{\bibinfo{person}{Kurt Thomas}, \bibinfo{person}{Patrick~Gage Kelley}, \bibinfo{person}{Sunny Consolvo}, \bibinfo{person}{Patrawat Samermit}, {and} \bibinfo{person}{Elie Bursztein}} (Eds.). \bibinfo{year}{2022}\natexlab{}.
\newblock \bibinfo{booktitle}{\emph{"It's common and a part of being a content creator'': Understanding How Creators Experience and Cope with Hate and Harassment Online}}.
\newblock


\bibitem[To et~al\mbox{.}(2021)]%
        {to21reducing}
\bibfield{author}{\bibinfo{person}{Alexandra To}, \bibinfo{person}{Hillary Carey}, \bibinfo{person}{Geoff Kaufman}, {and} \bibinfo{person}{Jessica Hammer}.} \bibinfo{year}{2021}\natexlab{}.
\newblock \showarticletitle{Reducing Uncertainty and Offering Comfort: Designing Technology for Coping with Interpersonal Racism}. In \bibinfo{booktitle}{\emph{Proceedings of the 2021 CHI Conference on Human Factors in Computing Systems}} (Yokohama, Japan) \emph{(\bibinfo{series}{CHI '21})}. \bibinfo{publisher}{Association for Computing Machinery}, \bibinfo{address}{New York, NY, USA}, Article \bibinfo{articleno}{398}, \bibinfo{numpages}{17}~pages.
\newblock
\showISBNx{9781450380966}
\urldef\tempurl%
\url{https://doi.org/10.1145/3411764.3445590}
\showDOI{\tempurl}


\bibitem[Venkit et~al\mbox{.}(2023)]%
        {venkit23automated}
\bibfield{author}{\bibinfo{person}{Pranav~Narayanan Venkit}, \bibinfo{person}{Mukund Srinath}, {and} \bibinfo{person}{Shomir Wilson}.} \bibinfo{year}{2023}\natexlab{}.
\newblock \showarticletitle{Automated Ableism: An Exploration of Explicit Disability Biases in Sentiment and Toxicity Analysis Models}.
\newblock \bibinfo{journal}{\emph{CoRR}}  \bibinfo{volume}{abs/2307.09209} (\bibinfo{year}{2023}).
\newblock
\urldef\tempurl%
\url{https://doi.org/10.48550/ARXIV.2307.09209}
\showDOI{\tempurl}
\showeprint[arXiv]{2307.09209}


\bibitem[Wallace et~al\mbox{.}(2013)]%
        {wallace13making}
\bibfield{author}{\bibinfo{person}{Jayne Wallace}, \bibinfo{person}{John McCarthy}, \bibinfo{person}{Peter~C. Wright}, {and} \bibinfo{person}{Patrick Olivier}.} \bibinfo{year}{2013}\natexlab{}.
\newblock \showarticletitle{Making design probes work}. In \bibinfo{booktitle}{\emph{Proceedings of the SIGCHI Conference on Human Factors in Computing Systems}} (Paris, France) \emph{(\bibinfo{series}{CHI '13})}. \bibinfo{publisher}{Association for Computing Machinery}, \bibinfo{address}{New York, NY, USA}, \bibinfo{pages}{3441–3450}.
\newblock
\showISBNx{9781450318990}
\urldef\tempurl%
\url{https://doi.org/10.1145/2470654.2466473}
\showDOI{\tempurl}


\bibitem[Wang and Ringland(2023)]%
        {wang23weaving}
\bibfield{author}{\bibinfo{person}{Yihe Wang} {and} \bibinfo{person}{Kathryn~E. Ringland}.} \bibinfo{year}{2023}\natexlab{}.
\newblock \showarticletitle{Weaving Autistic Voices on TikTok: Utilizing Co-Hashtag Networks for Netnography}. In \bibinfo{booktitle}{\emph{Companion Publication of the 2023 Conference on Computer Supported Cooperative Work and Social Computing}} (Minneapolis, MN, USA) \emph{(\bibinfo{series}{CSCW '23 Companion})}. \bibinfo{publisher}{Association for Computing Machinery}, \bibinfo{address}{New York, NY, USA}, \bibinfo{pages}{254–258}.
\newblock
\showISBNx{9798400701290}
\urldef\tempurl%
\url{https://doi.org/10.1145/3584931.3606995}
\showDOI{\tempurl}


\bibitem[Warner et~al\mbox{.}(2024)]%
        {warner24critical}
\bibfield{author}{\bibinfo{person}{Mark Warner}, \bibinfo{person}{Angelika Strohmayer}, \bibinfo{person}{Matthew Higgs}, {and} \bibinfo{person}{Lynne Coventry}.} \bibinfo{year}{2024}\natexlab{}.
\newblock \bibinfo{title}{A Critical Reflection on the Use of Toxicity Detection Algorithms in Proactive Content Moderation Systems}.
\newblock
\newblock
\showeprint[arxiv]{2401.10629}~[cs.HC]
\urldef\tempurl%
\url{https://arxiv.org/abs/2401.10629}
\showURL{%
\tempurl}


\bibitem[Wolbring(2008)]%
        {wolbring08politics}
\bibfield{author}{\bibinfo{person}{Gregor Wolbring}.} \bibinfo{year}{2008}\natexlab{}.
\newblock \showarticletitle{The Politics of Ableism}.
\newblock \bibinfo{journal}{\emph{Development}} \bibinfo{volume}{51}, \bibinfo{number}{2} (\bibinfo{year}{2008}), \bibinfo{pages}{252--258}.
\newblock
\urldef\tempurl%
\url{https://doi.org/10.1057/dev.2008.17}
\showDOI{\tempurl}


\bibitem[Zhang and SK(2023)]%
        {aws23}
\bibfield{author}{\bibinfo{person}{Lana Zhang} {and} \bibinfo{person}{Ravisha SK}.} \bibinfo{year}{2023}\natexlab{}.
\newblock \bibinfo{title}{Flag Harmful Content: Using Amazon Comprehend for Toxicity Detection}.
\newblock \bibinfo{howpublished}{\url{https://aws.amazon.com/blogs/machine-learning/flag-harmful-content-using-amazon-comprehend-toxicity-detection/}}.
\newblock
\newblock
\shownote{Accessed: September 10, 2024}.


\end{thebibliography}
\end{document}